\documentclass[fleqn,usenatbib]{mn2e}

\usepackage{newtxtext,newtxmath}

\usepackage[T1]{fontenc}

\DeclareRobustCommand{\VAN}[3]{#2}
\let\VANthebibliography\thebibliography
\def\thebibliography{\DeclareRobustCommand{\VAN}[3]{##3}\VANthebibliography}

\usepackage{graphicx}	
\usepackage{amsmath}	
\usepackage{amssymb}	

\usepackage{epsfig,amsmath,natbib}
\usepackage{color,subfigure}
\usepackage{natbibmnfix}
\usepackage{xcolor}

\usepackage{mathrsfs,amssymb,amstext}
\usepackage{ulem}
\usepackage{url}
\usepackage{bm}
\usepackage{adjustbox}

\def\refjnl#1{{\rm#1}}
\def\apjl{\refjnl{ApJ}}                
\def\apjs{\refjnl{ApJS}}               
\def\ao{\refjnl{Appl.~Opt.}}           
\def\aap{\refjnl{A\&A}}                

\def\be{\begin{equation}} 
\def\ee{\end{equation}} 
\def\ba{\begin{eqnarray}} 
\def\ea{\end{eqnarray}}

\def\gsim{\lower.5ex\hbox{\gtsima}} 
\def\lsim{\lower.5ex\hbox{\ltsima}} \def\gtsima{$\; \buildrel > \over 
\sim \;$} \def\ltsima{$\; \buildrel < \over \sim \;$} \def\prosima{$\; 
\buildrel \propto \over \sim \;$} \def\gsim{\lower.5ex\hbox{\gtsima}} 
\def\lsim{\lower.5ex\hbox{\ltsima}} 
\def\simgt{\lower.5ex\hbox{\gtsima}} 
\def\simlt{\lower.5ex\hbox{\ltsima}} 
\def\simpr{\lower.5ex\hbox{\prosima}} \def\la{\lsim} \def\ga{\gsim} 
  
\def\gtsima{$\; \buildrel > \over \sim \;$} 
\def\ltsima{$\; \buildrel < \over \sim \;$} 
\def\gsim{\lower.5ex\hbox{\gtsima}} 
\def\lsim{\lower.5ex\hbox{\ltsima}} 
\def\simgt{\lower.5ex\hbox{\gtsima}} 
\def\simlt{\lower.5ex\hbox{\ltsima}} 
\def\simpr{\lower.5ex\hbox{\prosima}} 
\def\la{\lsim} 
\def\ga{\gsim}

\def\E3{{\cal E}_{\rm g}^{III}}

\def\x12{x_{1/2}} 

\def\gtsima{$\; \buildrel > \over \sim \;$}
\def\ltsima{$\; \buildrel < \over \sim \;$}
\def\gsim{\lower.5ex\hbox{\gtsima}}
\def\lsim{\lower.5ex\hbox{\ltsima}}
\def\simgt{\lower.5ex\hbox{\gtsima}}
\def\simlt{\lower.5ex\hbox{\ltsima}}
\def\simpr{\lower.5ex\hbox{\prosima}}

\def\mean#1{\left< #1 \right>}

\title[Radio signals from early DCBHs]{Radio signals from early direct collapse black holes}
\author[Yue \& Ferrara]{
B. Yue$^{1}$ \& A. Ferrara$^{2,3}\thanks{E-mail: andrea.ferrara@sns.it}  $
\\
$^{1}$National Astronomical Observatories, Chinese Academy of Sciences, 20A, Datun Road, Chaoyang District, Beijing, 100101, China\\
$^{2}$Scuola Normale Superiore, Piazza dei Cavalieri 7, 56126 Pisa, Italy\\
$^{3}$Kavli Institute for the Physics and Mathematics of the Universe (WPI), University of Tokyo, Kashiwa 277-8583, Japan\\
}

\begin{document}
\label{firstpage}
\pagerange{\pageref{firstpage}--\pageref{lastpage}}
\maketitle

\begin{abstract}
 We explore the possibility to detect the continuum radio signal from direct collapse black holes (DCBHs) by upcoming radio telescopes such as the SKA and ngVLA, assuming that after formation they can launch and sustain powerful jets at the accretion stage. 
We assume that the high-$z$ DCBHs have similar jet properties as the observed radio-loud AGNs, then use a jet model to predict their radio flux detectability. 
If the jet power $P_{\rm jet}\gtrsim10^{42-43}$ erg s$^{-1}$, it can be detectable by SKA/ngVLA, depending on the jet inclination angle. 
Considering the relation between jet power and black hole mass and spin, generally, jetted DCBHs with  mass  $\gtrsim10^5~M_\odot$ can be detected.
For a total jetted DCBH number density of $\sim2.5\times10^{-3}$ Mpc$^{-3}$ at $z=10$, about 100 deg$^{-2}z^{-1}$ DCBHs are expected to be above the detection threshold of SKA1-mid (100 hours integration). 
If the jet ``blob'' emitting most of the radio signal is dense and highly relativistic, then the DCBH would only feebly emit in the SKA-low band, because of self-synchrotron absorption (SSA) and blueshift. 
Moreover, the free-free absorption in the DCBH envelope may further reduce the signal in the SKA-low band. 
Thus, combining SKA-low and SKA-mid observations might provide a potential tool to distinguish a DCBH from a normal star-forming galaxy.

\end{abstract}

\begin{keywords}
quasars: supermassive black holes -- radio continuum -- galaxies: jet
\end{keywords}



\section{Introduction}\label{introduction}
Supermassive black holes (SMBHs) routinely observed at redshift $z \simeq 6$ must have collected their mass, $\sim 10^{8-10}~M_\odot$, within $\sim 900$ million years from the Big Bang (e.g., \citealt{Wu2015,Banados2018}), starting from much smaller black hole seeds. The origin and nature of the seeds are still very uncertain, although several hypotheses have been proposed, ranging from stellar-mass black holes formed after the death of Pop III stars \citep{Heger2003}, to the collapse of nuclear star clusters, and even direct collapse black holes (DCBHs) that form directly in pristine halos in which H$_2$ cooling is suppressed (e.g., \citealt{Omukai2001,OhHaiman,Begelman06,Begelman08,Begelman10,Umeda2016}). 
Usually this is achieved by a super-critical radiation field that dissociates the H$_2$, and/or detaches the H$^-$ that is crucial for H$_2$ formation. The critical value $J_{21}$, which is the specific intensity of the radiation field in units of $10^{-21}$ erg s$^{-1}$cm$^{-2}$Hz$^{-1}$sr$^{-1}$ at 13.6 eV, is still not yet confirmed. It ranges between $\sim30-10^5$, depending on spectrum shape of the radiation source; how to model the self-shielding and hydrodynamical effects; and so on.
These alternatives are reviewed in \citet{Volonteri2010} and \citet{Latif2016}. If the seed is a stellar-mass black hole, then it must always keep accreting material at the highest rate (Eddington limit) throughout the Hubble time. This is quite unlikely \citep{Valiante2016}. Numerical simulations show that, at the early growth stage, radiative feedback is efficient and the accretion rate is therefore limited to  $\simeq 10^{-5}-10^{-3}$ of the Eddington limit \citep{Alvarez2009}.  Also \citet{Smith2018} found that for 15000 stellar-mass black holes hosted by minihalos in simulations, the growth is negligible. 

The DCBH scenario, at least in principle, solves this tension. Such kind of black holes could be as massive as $\sim10^{4-6}~M_\odot$ at birth, when they find themselves surrounded by pristine gas with high temperature and density. This is an ideal situation to sustain the high accretion rate required by SMBH growth \citep{Paucci2015b,Latif2020, Regan2020}.

DCBHs have not been detected until now, either because they are too faint and/or too rare. Potentially promising detections techniques have been proposed in the literature. They include, among others, the cosmic infrared background \citep{Yue2013b}, multi-color sampling of the Spectral Energy Distribution (SED) combined with X-ray surveys \citep{Pacucci2016}, a specific Ly$\alpha$ signature \citep{Dijkstra2016b}, and the neutral hydrogen $\lambda =3$ cm maser line \citep{Dijkstra2016a}. These methods are summarized in \citet{Dijkstra2019}.

Here we introduce a new strategy based on the detection of the radio continuum signal from early DCBHs. 
We consider a DCBH population featuring a powerful jet, similar to what observed in the radio-loud AGN (or blazars, if the jet beam happens to point towards the observer). Such a DCBH must be at the  accretion stage, the collapse stage before the DCBH formation, including the lifetime of possible supermassive star, is much shorter (e.g. \citealt{Begelman2010}).    
Launching the jet relies on the presence of a very strong magnetic field close to the black hole surface.  It is found that indeed the magnetic field amplification mechanism works during the accretion process around a DCBH \citep{Latif2014_magnetic}. And, \citet{Sun2017} simulated the collapse of a supermassive star (the progenitor of a DCBH), showing that the end result of such process is a high spin black hole with an incipient jet. Moreover, \citet{Matsumoto2015,Matsumoto2016} showed that the jet can indeed break out the stellar envelope of the supermassive star. Based on these theoretical predictions, we consider the possible existence of jetted DCBHs at high redshifts, and use this assumption in the paper.

For jetted DCBHs, we first use an empirical jet power-radio power relation to estimate the strength of radio flux at the bands of upcoming radio telescopes SKA-low, SKA-mid and ngVLA in Sec. \ref{sec:radio-flux}; then, we combine the spectral radiation model (Sec. \ref{sec:jet-radiation})  and the jet total power  model (Sec. \ref{sec:jet-power}) to make detailed predictions on the expected radio flux and the DCBH detectability. 
In Sec. \ref{sec:envelope} we show the free-free absorption of the DCBH envelope. 
We give summary in Sec. \ref{sec:summary}.

Appendix \ref{sec:syn} gives the synchrotron radiation formulae.
In Appendix \ref{sec:BH-spin} we estimate the angular momentum   supply in the accretion disk, to see if it is enough to support  a black hole with high-spin.
Appendix \ref{sec:radio-quiet} contains the results for radio-quiet DCBHs. For radio-quiet DCBHs, we follow \citet{Yue2013b} who developed a radiation model. The primary radiation includes the contribution from a disk, a hot corona and a reflected component, it is re-processed by the envelope material. In \citet{Yue2013b} the radio emission is mainly produced by free-free radiation; however, we add some synchrotron contribution that is not considered by the previous work.  
In Appendix \ref{sec:jet-prop} we estimate the propagation of a jet in different DCBH envelopes, to see in which case the jet can break out the envelope. If a jet can break out the envelope then it does not suffer the free-free absorption.

\section{Jetted DCBHs}

\subsection{Radio flux from the jet}\label{sec:radio-flux}
 
We start by obtaining a rough estimate of the jet radio flux using the known empirical relation between the jet power and the radio power. The relation is approximated by a power-law form and the coefficients obtained from a number of observations are consistent with each other (e.g., \citealt{Birzan2008,Cavagnolo2010,OSullivan2011,Daly2012}). 
However, \citet{Godfrey2016} pointed out that this apparent consistence is due to the bias in measuring jet power of samples that span large distance range, and found different relations for different sample classifications.

\begin{figure}
\centering{
\subfigure{\includegraphics[width=0.45\textwidth]{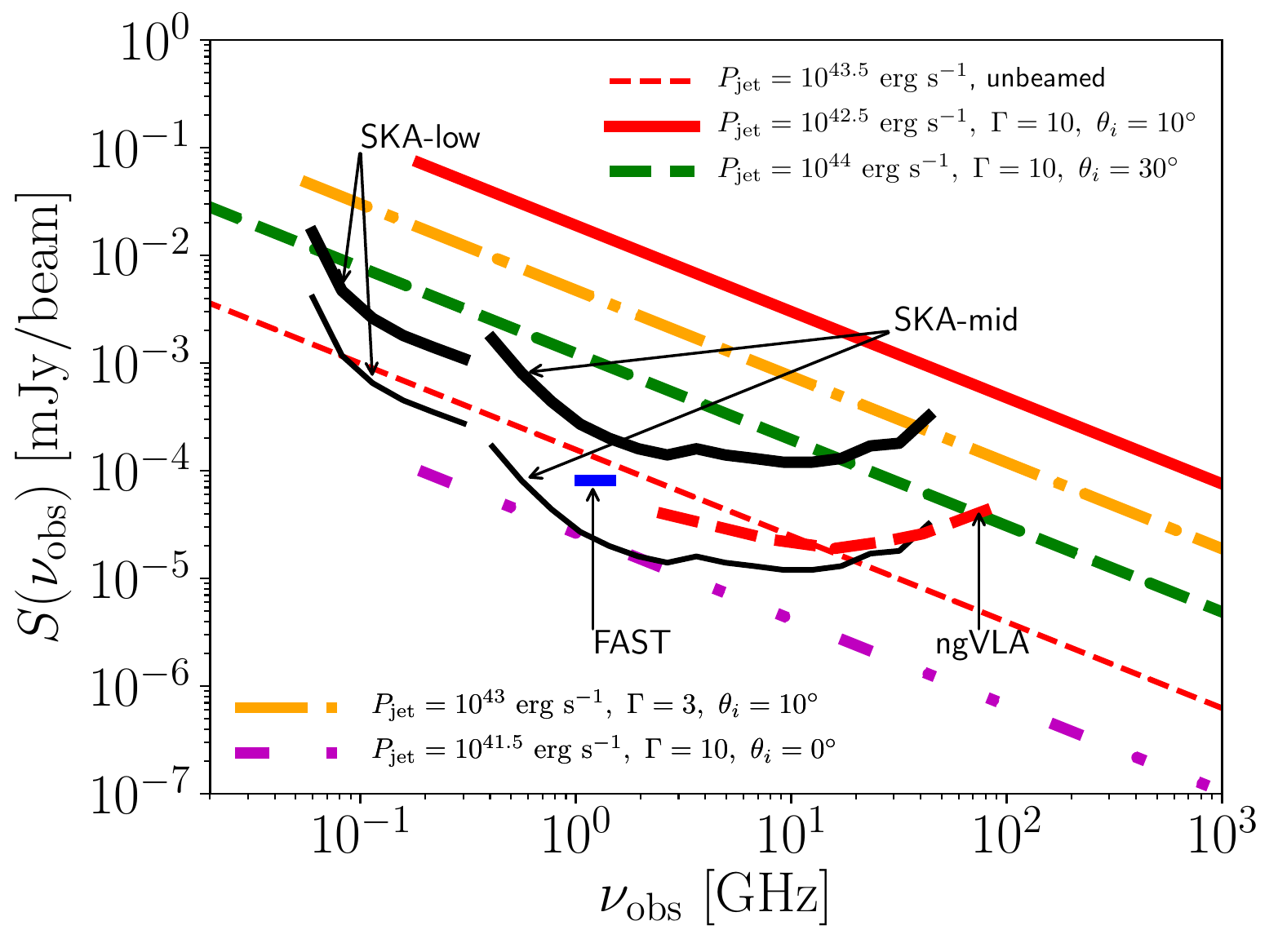}}
\caption{Observed radio flux from a jetted DCBH, for different jet powers, Lorentz factors and inclination angles, as indicated in the label. Sensitivities of different telescopes are given. For all sensitivities the integration time is 100 hours, and bandwidth is given in the text. 
}
\label{fig:S_obs}
}
\end{figure}

Nevertheless, we adopt the relation for radio luminosity at 1.4 GHz in \citet{Birzan2008}, which is actually close to that is found in \citet{Godfrey2016},
 \begin{equation}
 \log \left(\frac{ P_{\rm jet}}{10^{42}\rm erg~s^{-1}}\right)=p_0 \log\left( \frac{L_{1.4}} {10^{24}\rm W~Hz^{-1}}\right)+p_1, 
 \label{eq:P_jet-L_radio} 
 \end{equation}
where $L_{1.4}$ is the luminosity at 1.4 GHz (rest-frame), and $p_0=0.35$, $p_1=1.85$. The observed flux is then
\begin{equation}
S(\nu_{\rm obs})= \frac{L_\nu(\nu')(1+z)}{4\pi d_L^2}
\end{equation}
if the energy is isotropically distributed in space, where $\nu'=\nu_{\rm obs}(1+z)$ and $d_L$ is the luminosity distance up to $z$. Assuming that $L_\nu\propto \nu^{-0.8}$, for $z=10$ and $P_{\rm jet}=10^{43.5}$ erg s$^{-1}$, the observed flux is shown by a thin dashed line in Fig. \ref{fig:S_obs}. Using \citet{Cavagnolo2010} and  \citet{OSullivan2011} relations gives similar flux for $P_{\rm jet}=10^{43.5}$ erg s$^{-1}$. However since their $p_0\sim0.7$,  they will give much lower flux for  $P_{\rm jet}\gtrsim 10^{43.5}$ erg s$^{-1}$, and much higher flux for $P_{\rm jet}\lesssim 10^{43.5}$ erg s$^{-1}$.

For a highly relativistic jet, the beaming effect can boost/reduce the signal \citep{Ghisellini2000},
\begin{equation}
S(\nu_{\rm obs})=\delta^3_{\rm D}\frac{L'_\nu(\nu')(1+z)}{4\pi d_L^2},
\end{equation}
here $L'_\nu$ is the radio luminosity at the jet comoving frame. 
The Doppler factor can be written as   
\begin{equation}
\delta_{\rm D}=\frac{1}{\Gamma (1-\beta_j \cos\theta_i)},
\end{equation}
where $\Gamma$ is the Lorentz factor, and $\beta_j$ is the jet velocity in units of the light speed; $\theta_i$ is the inclination angle of the jet axis with respect to the line-of-sight.
In this case the relation between the observed frequency $\nu_{\rm obs}$, and the jet comoving frame frequency $\nu'$ is 
 \begin{equation}
\nu_{\rm obs}=\frac{\nu'\delta_{\rm D}}{1+z}.
\end{equation}

However, the radio luminosity in Eq. (\ref{eq:P_jet-L_radio}) is
derived by assuming that the radio power is isotropically distributed,
without beaming correction. Here we treat  $L_{1.4}$ as in the jet
comoving frame, with the warning that a beaming-induced bias might be
present in the observed samples.

In Fig. \ref{fig:S_obs} we plot the observed radio flux vs. $\nu_{\rm obs}$ for different jet power, inclination angle and Lorentz factor. 
We cut the intrinsic luminosity below $\nu'=0.1$ GHz, so that the redshift/blueshift effects can be observed from the cut-frequency of the observed flux.
We also plot the sensitivity for SKA1-low/SKA2-low and SKA1-mid/SKA2-mid for fractional bandwidth 0.3 \citep{Braun_2019_SKA1}.
We assume the SKA2-low has sensitivity four times higher than SKA1-low, while SKA2-mid has sensitivity ten times higher than SKA1-mid \citep{Braun2014}. We also plot the sensitivity of the ngVLA, see \citet{ngVLA2018} where the bandwidth is also given, and the sensitivity of FAST telescope for a 400 MHz bandwidth \citep{Li2018RAA}. For all telescopes we adopt 100 hours integration time. Note that FAST has low angular resolution, with a beam size $\sim 2.5'$ at 1.4 GHz. So one must take into account the confusion noise as well \citep{Condon2012} when observing distant objects. 

From these results in Fig. \ref{fig:S_obs} we conclude that if a $z \simeq 10$ DCBH can launch a jet with $ P_{\rm jet}\sim10^{42}$ erg s$^{-1}$, and if the scaling relation Eq. (\ref{eq:P_jet-L_radio}) holds, it will be possible to detect it at radio frequencies with both SKA and ngVLA in about $100$ hours of observation, depending on the inclination angle $\theta_i$ and $\Gamma$. If the jet is powerful enough, lower $\Gamma$ values improve detection chances as the emission is less beamed. 
However, considering the large scatters in observations on the $P_{\rm jet}-L_{1.4}$ relation, and possible bias in measuring the jet power, the above is just to present a very rough estimate.

\subsection{Jet radiation spectrum}\label{sec:jet-radiation}   

 To refine our previous analysis, we need to discuss more in depth the details of the jet synchrotron emission. Luckily, a considerable amount of works are present in the literature. We will follow, in particular, the study by \citet[][GT09]{Ghisellini2009} which concentrates on blazars, see also \citet{Ghisellini2010} and \citet{Gardner2018}.  Their code is also publicly available\footnote{\url{https://starchild.gsfc.nasa.gov/xanadu/xspec/models/jet.html}}. We will use GT09 results to assess whether the physical parameters required for a jet to produce an observed radio flux similar to Sec. \ref{sec:radio-flux} are reasonable. Below we briefly summarize the model main features, referring the reader to the original papers for details.

GT09 is a one-zone leptonic jet model. Most of the jet radiation originates from a spherical ``blob'' that forms when the jet dissipates. The blob contains relativistic electrons and has a bulk motion along the jet propagation direction. 
Given the energy spectrum of injected relativistic electrons in the blob, the magnetic field $B$, the Lorentz factor of the bulk motion $\Gamma$, the physical size of the blob $r_{\rm blob}$, and the power of the injected relativistic electrons in jet comoving frame $P'_{\rm i}$, the model returns the power of the jet in each component: relativistic electrons, $P_e$; protons $P_p$; magnetic field $P_B$; and radiation $P_r$. 
The jet power is the sum of these four components: $P_{\rm jet}=P_e+P_p+P_B+P_r$.
The radiation power $P_r$ is mainly produced through two mechanisms: synchrotron and Compton scattering.
The observed radio flux at SKA and ngVLA band is from  the synchrotron part.
Finally, it gives the observed radiation spectrum of the jet and of the whole system. 

Regarding the jet properties, \citet{Ghisellini2010} obtained the properties of 85 blazars with $z\lesssim3$, including Flat Spectrum Radio Quasars (FSRQs) and BL Lacs,  by fitting their observed SEDs using the GT09 model. They found that the mean physical radius of the blob is $r_{\rm blob}\simeq 50~r_S$, where $r_S$ is Schwarzschild radius of the central black hole. Blobs have a bulk motion with a mean Lorentz factor $\Gamma\sim10$, and mean magnetic field $B\sim1$ G. The energy spectrum of injected relativistic electrons in the blob is assumed to be a broken power law, with mean break Lorentz factor $\gamma_{\rm break}\sim300$, and mean maximum Lorentz factor $\gamma_{\rm max}\sim3\times10^3$ for FSRQs; and $\gamma_{\rm break}\sim1.5\times10^4$,  $\gamma_{\rm max}\sim8\times10^5$ for BL Lacs. The slopes of the energy spectrum are uncertain, but typically below the break point the spectral slope is $s_1\sim1$ and above it is $s_2\sim3$.  They also found that the ratio $P_{\rm r}/P_{\rm jet}$ is in the range $\sim0.01-0.5$, and that the jet power is generally larger than the disk luminosity. \citet{Ghisellini2015} found similar properties for several $z\gtrsim5$ blazars: $r_{\rm blob}\sim50~r_S$, $B\sim1$ G, $\Gamma\sim10$, $\gamma_{\rm break}\sim70-300$, $\gamma_{\rm max}\sim3000$, and $P_{\rm r}/P_{\rm jet}\sim0.02-0.13$. \citet{Paliya2017,Paliya2019,Paliya2020} analyzed a sample of   blazars and found $r_{\rm blob}\sim50-200~r_S$, $B\sim1$ G, $\Gamma\sim10$, the spectrum index of the electrons  distribution    below $\gamma_{\rm break}$  is $\sim2$ where $\gamma_{\rm break}\sim100$, and $P_{\rm r}/P_{\rm jet}\sim0.02-0.13$.

\begin{table*}
\caption{The jet parameters for the observed flux shown in Fig. \ref{fig:Fobs_jet-GT09}}.
\begin{center}
\begin{tabular}{|c|c|c|c|c|c|c|c|c|c|c|c|c|c|c|}
\hline\hline
$^{(0)}$No. & $^{(1)}\Gamma$ &     $^{(2)}r_{\rm blob}$&  $^{(3)}B$&     $^{(4)}\gamma_{\rm min}$ & $^{(5)}\gamma_{\rm break}$ & $^{(6)}\gamma_{\rm max}$ & $^{(7)}s_1$ & $^{(8)}s_2$    \\
 &\text{--} &      [$r_{\rm S}$]&[G] & \text{--}  &\text{--}&\text{--}&\text{--}&\text{--}   \\ 
 
\hline
   
 1&10 &  $50$  &    1  &   1& $3\times10^2$ & $5\times10^3$ & 1.0 & 3.0  \\
 
 2&3 &  $50$  &    1  &   1& $3\times10^2$ & $5\times10^3$ & 1.0 & 3.0  \\

3&10 &  $50$  &    10  &  1& $3\times10^2$ & $5\times10^3$ & 1.0 & 3.0  \\  

4&10 &  $2000$  &    1  &   1& $1\times10^4$ & $1\times10^6$ & 2.0 & 3.0  \\

\hline
\end{tabular}
\end{center}
\flushleft
Col. (0): jet No.;
Col. (1): Lorentz factor of the blob bulk motion;
Col. (2): physical radius of the blob; 
Col. (3): magnetic field in the blob;
Col. (4): minimum Lorentz factor of relativistic electrons;
Col. (5): Lorentz factor at the break;
Col. (6): maximum Lorentz factor;
Col. (7): spectrum index of the electron distribution between $\gamma_{\rm min}$ and $\gamma_{\rm break}$;
Col. (8): spectrum index of the electron distribution between $\gamma_{\rm break}$ and $\gamma_{\rm max}$;
\label{tab:jet-GT09}
\end{table*}

Motivated by the above observations, we adopt some jet parameter sets
for the following investigation, see the jet 1, 2 and 3 listed in
Tab. \ref{tab:jet-GT09}. These parameters are typical ones, i.e. not
rare among the above observed  samples. We assume the black hole mass
$M_{\rm BH}=10^6~M_\odot$ (Eddington luminosity is $L_E = 1.4\times10^{44}$ erg s$^{-1}$) at redshift $z=10$, then calculate the synchrotron radiation from the jet of a high-$z$ DCBH with these parameters. We ignore the effects of other parameters that are less relevant to synchrotron. GT09 uses a $\delta$-function to approximate synchrotron  from electrons with a given Lorentz factor. We slightly improve their code to allow for a more realistic electron energy distribution, as described in Appendix \ref{sec:syn}. 

\begin{figure*}
\centering{
\subfigure{\includegraphics[width=0.85\textwidth]{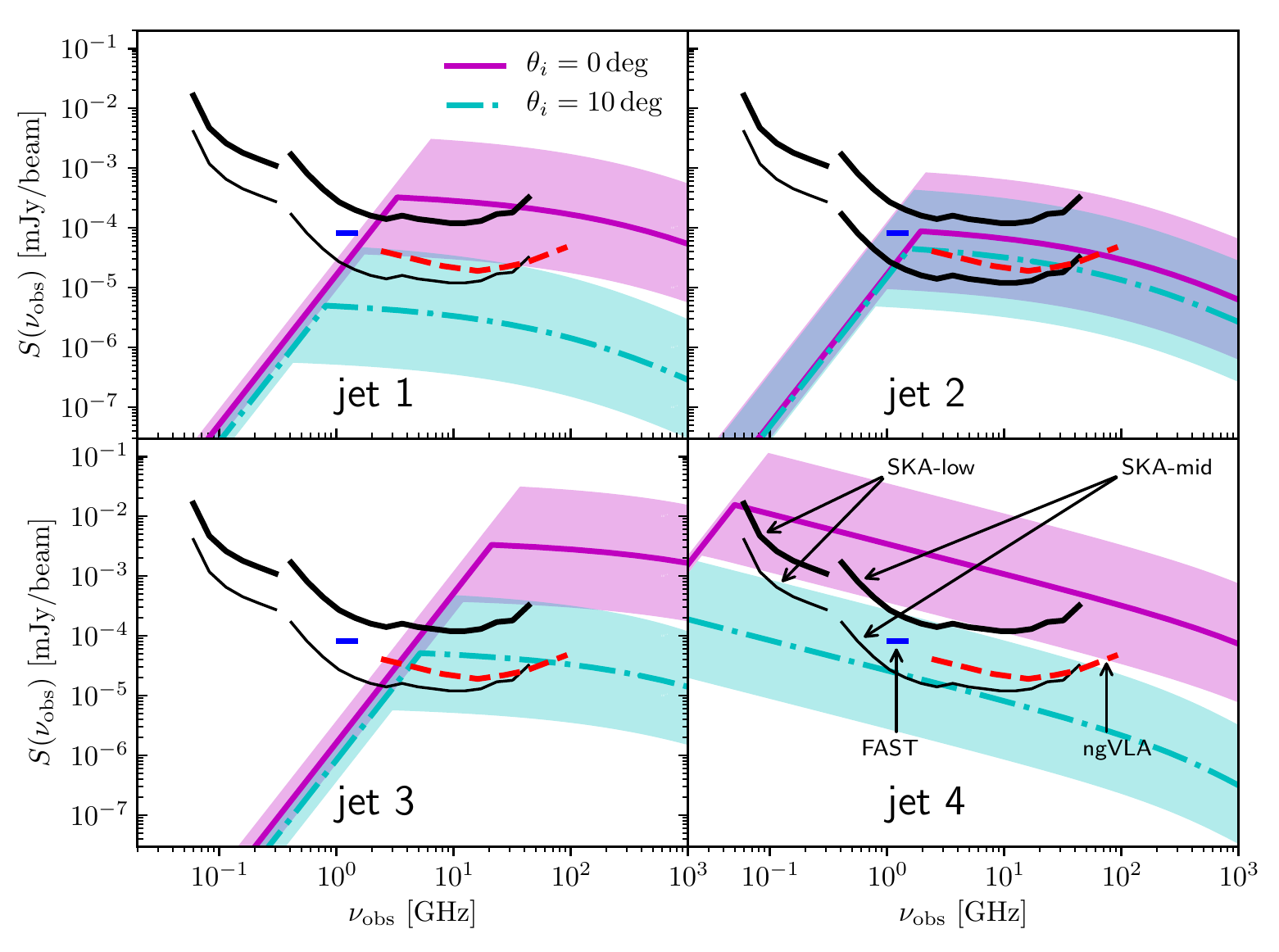}}
\caption{Observed radio flux from a jetted DCBH at $z=10$ with emission parameters listed in Tab. \ref{tab:jet-GT09}. Sensitivities of various radio telescopes are the same as in Fig. \ref{fig:S_obs}.
In each panel we show the results for inclination angle 0 and 10 degree (solid and dashed-dotted lines), and jet power from 0.1 to 10 Eddington luminosity (shaded regions around each line). The lines are for jet power equal to the Eddington luminosity.
}
\label{fig:Fobs_jet-GT09}
}
\end{figure*}

We plot the observed radio spectrum in
Fig. \ref{fig:Fobs_jet-GT09}. For each model, we plot the curve for a
total jet power equal to the Eddington luminosity. Around the curve,
the filled color shows the range corresponding the jet power from 0.1 $L_E$
to 10 $L_E$. We show the results for inclination angles of  0 and 10
degrees, respectively. For large inclinations, the flux is further
reduced. Since $r_{\rm blob}$ is given in units of Schwarzschild
radius and the jet power is given in units of $L_E$, their actual physical values are scaled to match DCBHs. 

For jet 1, 2 and 3 in Tab. \ref{tab:jet-GT09}, we confirm that the jet
emission is detectable by SKA-mid or ngVLA as long as the jet power is
$\gtrsim0.1$ of the Eddington luminosity, and the jet has a small
inclination angle. In particular, $\Gamma$, $r_{\rm blob}$ and
$P'_{\rm i}$ are important for generating strong synchrotron emission, while the inclination angle crucially determines whether such emission is detectable.  
For the SKA-low, the signal is not detectable. This is because if the blob is luminous and has small size as jet 1, 2 and 3, then it must be dense. The synchrotron self-absorption (SSA) would suppress low-frequency radiation, making the signal hard to detect by SKA-low. For this reason we consider a large blob size 2000 $r_S$, and also increase the Lorentz factors of relativistic electrons therein, see jet 4 in Tab. \ref{tab:jet-GT09}. 
A large blob size up to several thousand Schwarzschild radius is
physically conceivable.  For example, M87 has a jet that starts to
become conical when the width is $\sim2000$ $r_S$, as observed by
\citet{Asada2012} and supported by numerical models
\citep{Nakamura2018}.
We plot these results in Fig. \ref{fig:Fobs_jet-GT09} as well. We find
that in this case, as long as the inclination angle is not too small,
then the SSA effect is weaker, and the signal is detectable even at frequencies as low as 100 MHz. 
Quite interestingly, the SSA frequency-dependent signal suppression effect offers, at least in a certain parameter range, a direct way to discriminate between a radio source powered by a jetted DCBH with respect to, e.g., normal star-forming galaxies in which SSA should be essentially negligible.

\subsection{The envelope surrounding DCBH}\label{sec:envelope}

DCBH may be embedded into a warm and ionized envelope whose radius is
much larger than the Bondi radius; its column number density can reach
up to $\sim10^{25}$ cm$^{-2}$. If the jet can break out the envelope
and the radio blob forms outside it, then radio signal is not
influenced. However, in the opposite case the radio blob forms within
the envelope and the radio signal would be absorbed due to the
free-free absorption. Here we estimate the absorption level. The envelope also produces free-free emission. This is the main radio signal in case of no synchrotron radiation, see Appendix \ref{sec:radio-quiet}. 

 From numerical simulations, the radial density profile   for the envelope is \citep{Latif2013}
 \begin{equation}
n_{\rm a}(r)=n_0\frac{1}{1+(r/r_0)^\alpha},
\label{eq:n_a}
\end{equation}
where $n_0$, $r_0$ and $\alpha$ are free parameters. Simulations show that $\alpha=2$ and we adopt this value in this paper. We assume that both hydrogen and free electrons follow this profile. This profile evolves as the accretion proceeds. A final steady profile is likely $r_0\sim$ Bondi radius $\sim1$ pc \citep{Pacucci2015}.

The free-free absorption optical depth is \citep{Draine2011}
\begin{equation}
\tau_{\rm ff}(\nu)=0.05468\, T_e^{-1.5}\,\nu_9^{-2}\,{\rm EM}\, g_{\rm ff}, 
\label{eq:free-free}
\end{equation}
where $\nu_9$ is the frequency in units $10^9$ Hz and $T_e$ is the electron temperature;
the emission measure (in units of pc cm$^{-6}$)
\begin{equation}
{\rm EM}=\int_{z_{\rm blob}} n^2_a(r) dr,
\end{equation}
where $z_{\rm blob}$ is the distance of the dissipation blob from the central black hole;
and the Gaunt factor
\begin{align}
g_{\rm ff}(\nu)=\ln\left\{\exp\left[5.960-\frac{\sqrt{3}}{\pi}\ln\left(  \nu_9\, T_4^{-1.5} \right)\right]+\exp(1)\right\},
\label{eq:free-free}
\end{align}
where $T_4=T_e/10^4$K is the electron temperature in units $10^4$ K.
 
\begin{figure}
\centering{
\subfigure{\includegraphics[width=0.5\textwidth]{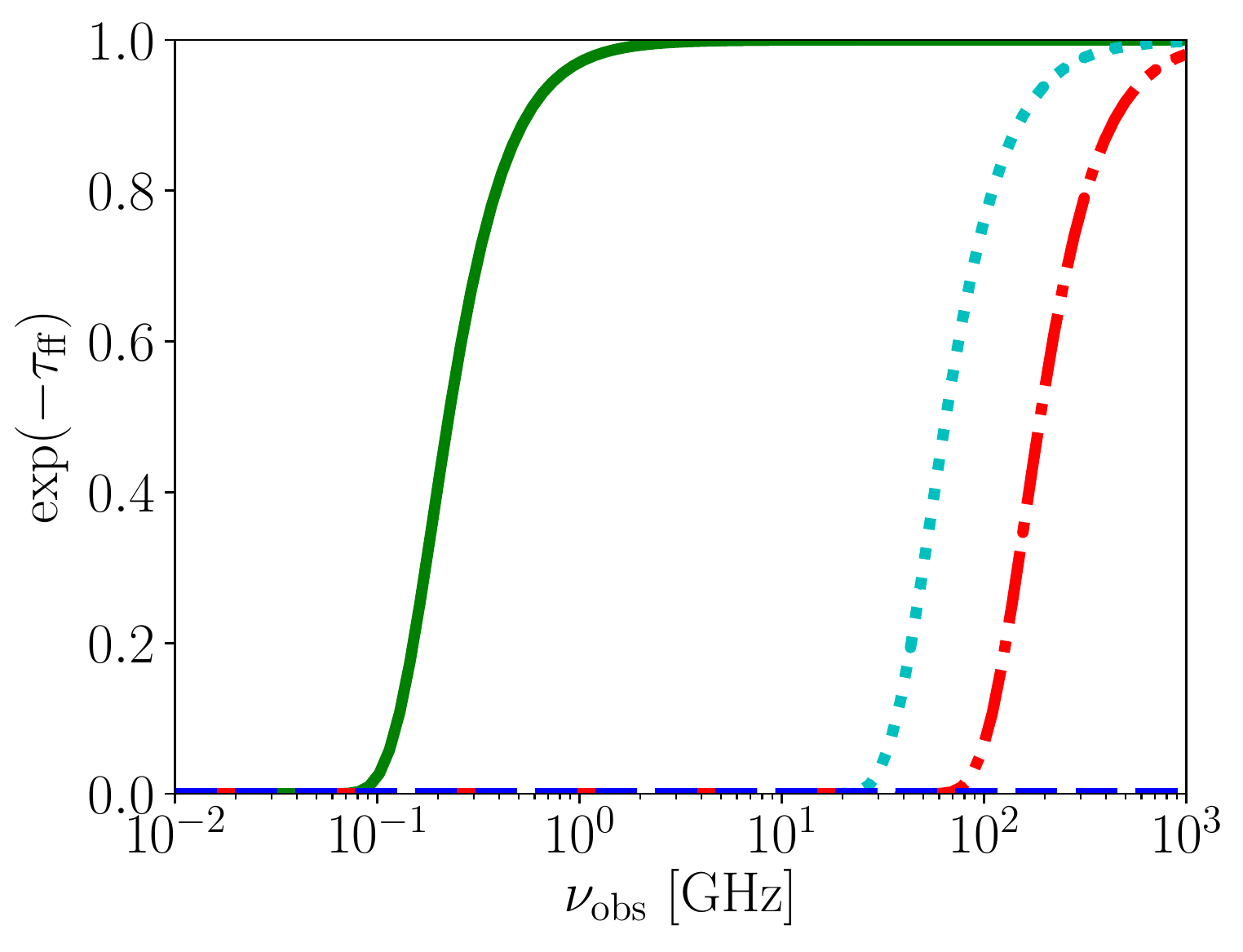}}
\subfigure{\includegraphics[width=0.5\textwidth]{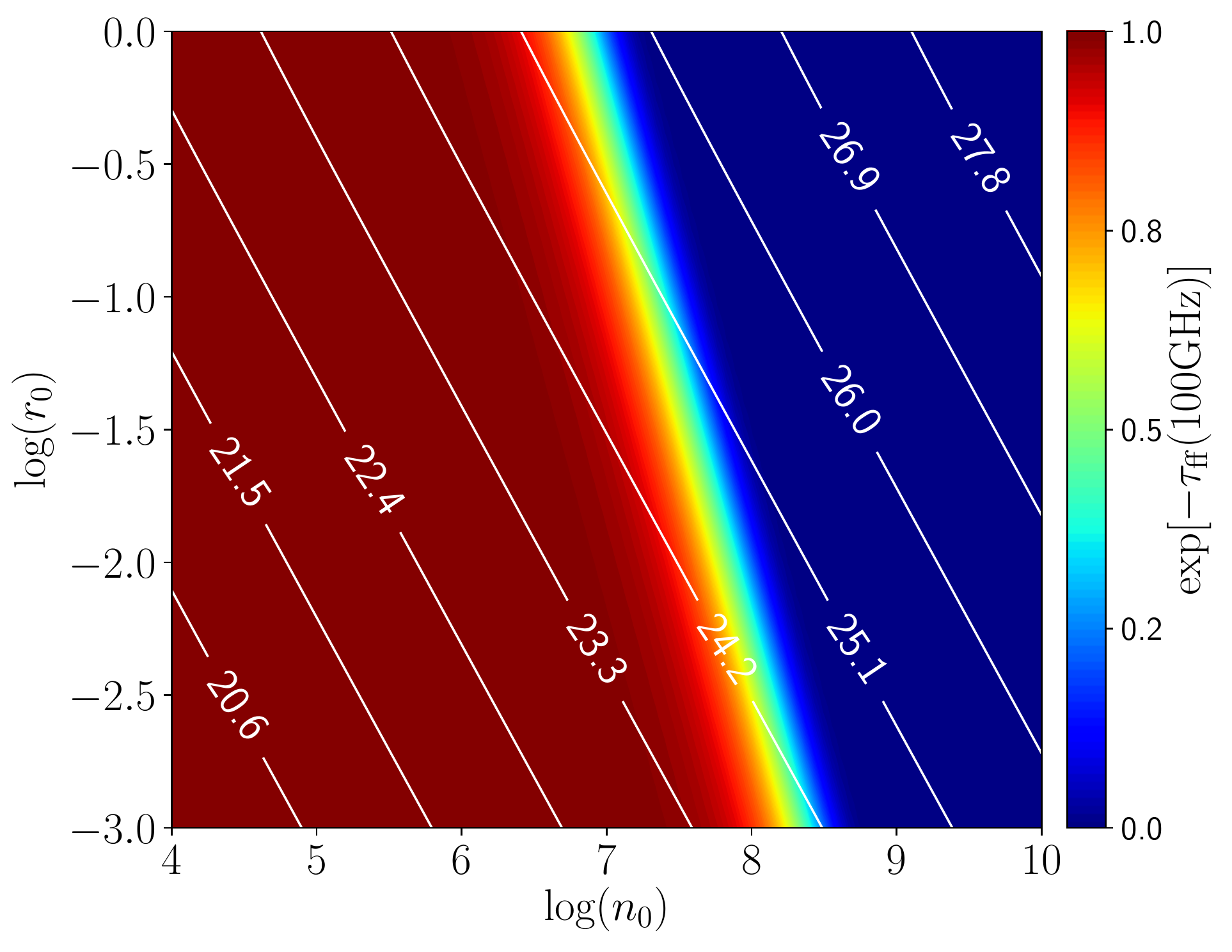}}
\caption{{\it Top:} The $\exp(-\tau_{\rm ff})$ curve for different envelope profiles. In the panel from left to right, the curves correspond to  $n_0=10^4$ cm$^{-3}$, $r_0=1$ pc;  $n_0=10^6$ cm$^{-3}$, $r_0=10^{-1}$ pc;  $n_0=10^8$ cm$^{-3}$, $r_0=10^{-2}$ pc and 
$n_0=10^{10}$ cm$^{-3}$, $r_0=10^{-3}$ pc respectively. 
{\it Bottom:} The $\exp(-\tau_{\rm ff})$ at $\nu_{\rm obs}=100$ GHz as a function of $n_0$ and $r_0$. We mark the $\log(N_{\rm H})$ as well by lines.
\label{fig:tau_ff}
}
}
\end{figure}

We plot the curve $\exp[-\tau_{\rm ff}(\nu)]$ in top panel of
Fig. \ref{fig:tau_ff} for some $n_0$ and $r_0$. We always assume
$z_{\rm blob}=r_0$, implicitly assuming that this is the maximum
propagation distance of the jet trapped in the envelope. Indeed, we
see that if the jet blob is within the envelope, free-free absorption
significantly reduces the radio signal. This represents a severe problem for DCBH detection  by SKA or ngVLA because the radio signal below $\sim100$ GHz would be absorbed ($n_0\gtrsim10^8$ cm$^{-2}$ and/or $r_0\gtrsim 0.1$ pc).

In the bottom panel of Fig. \ref{fig:tau_ff}  we plot the $\exp(-\tau_{\rm ff})$ at $\nu_{\rm obs}=100$ GHz and the log of column number density, $\log (N_{\rm H})$, as a function of $n_0$ and $r_0$. From the Fig. \ref{fig:tau_ff}, to make sure that the radio signal at $100$ GHz is  less absorbed, the column number density of the envelope should be $N_{\rm H}\lesssim 10^{24}-10^{25}$ cm$^{-2}$. We also checked that for signal detectable at 17.09 GHz, the column density should be $N_{\rm H}\lesssim10^{23}-10^{24}$ cm$^{-2}$. These are all reasonable values for typical DCBHs. 

Using the jet propagation model theory in \citet{Bromberg2011} (see
Appendix \ref{sec:jet-prop}), we estimate whether the jet can break
out of the envelope. We find that, for shallow density profile
($\alpha\sim2$), this does not occur unless the column density is very
small. However, if the envelope has a  steeper density profile,
i.e. $\alpha \sim 3$, the jet can easily break out even if the column
density is large.

Because of the uncertainties on $n_0$ and $r_0$,  we will not consider
absorption by the  envelope when estimating the radio
detectability. However, from now on we will concentrate on the SKA-mid
and ngVLA bands that are less influenced by such effect.

\subsection{Jet power supply}\label{sec:jet-power}

In Sec. \ref{sec:radio-flux}  and \ref{sec:jet-radiation}, we note that a radio detection requires $P_{\rm jet} \gtrsim 10^{42-43}$ erg s$^{-1}$, so the next key question is to verify whether such power can be physically extracted from the rotational energy of the system.   

Jets are usually described in the framework of the \citet[][BZ]{BZ77} or \citet[][BP]{BP82} popular models. In the  BZ model, the power of jet is originally from the spin of black hole, while in the BP model the jet is powered by the spin of the accretion disk. Here we adopt the BZ model because it is considered more suitable for relativistic jets \citep{Yuan2014,McKinney2005}, and it is supported by some observations, e.g. \citealt{Ghisellini2014} (for a different interpretation, however, see, e.g. \citealt{Fender2010}). 

In the BZ model,  the jet power originates from the black hole spin energy and it is extracted via a magnetic field. The relation between the jet power, black hole mass, spin and the magnetic flux is
\begin{equation}
P_{\rm jet}=\frac{\kappa}{4\pi c}\Phi_{\rm BH}^2\Omega_{\rm H}^2f(\Omega_{\rm H}),
\label{eq:P_BZ}
\end{equation}
where $f(\Omega_{\rm H})\approx1+1.38(\Omega_{\rm H}r_g/c)^2-9.2(\Omega_{\rm H}r_g/c)^4$ (\citealt{Tchekhovskoy2010,Tchekhovskoy2011,Tchekhovskoy2012,Tchekhovskoy2012b}); $\kappa\approx0.05$; $\Omega_{\rm H}=ac/2r_{\rm H}$ is the angular frequency. Furthermore, $r_{\rm H}=r_g[1+(1-a^2)^{1/2}]$, $r_g=GM_{\rm BH}/c^2$, $a=Jc/GM_{\rm BH}^2$  is the dimensionless spin parameter; $\Phi_{\rm  BH}$ is the magnetic flux threading the black hole. To launch a powerful jet, a rather strong magnetic flux is necessary. The magnetically arrested disc (MAD) model predicts that the poloidal magnetic flux can reach a maximum saturation value $\Phi_{\rm MAD}\approx 50 (\dot{M}_{\rm BH}r^2_gc)^{1/2}$ \citep{Narayan2003,Tchekhovskoy2011}. With this maximum magnetic flux, assuming radiative efficiency $\epsilon\gtrsim0.1$, the jet power could be larger than the Eddington luminosity when $a\gtrsim0.4$.
When $a\gtrsim0.9$  it could be even larger than the net inward power $P_{\rm in}=\dot{M}_{\rm BH}c^2=1.26\times10^{38}M_{\rm BH}(1-\epsilon)/\epsilon$ erg s$^{-1}$. So if such a jet lasts for a long time, the accretion must be radiatively inefficient. This was already pointed out by \citet{Tchekhovskoy2010}.

From the above discussion, it is clear that to supply the jet power a high-spin central black hole is required.
In Appendix \ref{sec:BH-spin}, we estimate the angular momentum in the dark matter host halo of a DCBH, and in the innermost part of the accretion disk. We show there that these reservoirs contain enough angular momentum to support a high-spin black hole.

As a final comment, we warn that it is unclear whether the poloidal magnetic flux can reach the MAD saturation level. In fact, this large field amplification might be prevented by, e.g. the magnetic flux Eddington limit set by equipartition between B-field,  and radiation field energy density near the BH surface \citep{Dermer2008}. In spite of these uncertainties, we will assume that the MAD value can be nevertheless achieved, as an optimistic forecast.

\section{Signal detectability}

For a given black hole, the BZ jet model provides the total jet power $P_{\rm jet}$, while the GT09 model gives the power distribution in each component: relativistic electrons $P_e$, protons $P_p$, magnetic field $P_B$  and radiation $P_r$. Hence, $P_{\rm jet}=P_e+P_p+P_B+P_r$ must hold. In the GT09 model, in addition to the energy spectrum of relativistic electrons in the blob, other input parameters that are most relevant to the synchrotron radiation are:  $B$,  $\Gamma$,  $r_{\rm blob}$,  and the total injected power of relativistic electrons in comoving frame, $P'_{\rm i}$ \citep{Ghisellini2010b,Gardner2014}. Since \citet{Ghisellini2010} found that $B$, $\Gamma$ and $r_{\rm blob}$ depend only weakly on $P_{\rm jet}$, also to  simplify the treatment,  in this work we assume they are all independent of  $P_{\rm jet}$. 

 Given the black hole mass and spin, we first get its jet power $P_{\rm jet}$ from Eq. (\ref{eq:P_BZ}). In the GT09 model, given the energy spectrum of relativistic electrons in the blob, and $B$, $\Gamma$, $r_{\rm blob}$ and $P'_{\rm i}$, one can get the $P_e, P_p, P_B,$ and $P_r$. 
However, by letting $P_e+P_p+P_B+P_r=P_{\rm jet}$ we can reduce the freedom by one, i.e., by iteration we derive a  $P'_{\rm i}$ for which $P_e+P_p+P_B+P_r=P_{\rm jet}$ holds. We then obtain the final observed synchrotron flux of this black hole for different $\theta_i$.

We further assume that the jet pointing direction is an uniform random distribution. Then the cumulative probability to observe a DCBH 
(given the mass, spin, redshift, and jet parameters) with flux $> S$ writes
\begin{equation}
P_a(>S|M_{\rm BH},a,B,\Gamma,r_{\rm blob}) =  \frac{ \{1-{\rm cos}[\theta_{i,S}(M_{\rm BH},a,B,\Gamma,r_{\rm blob})]\} }{2},
\label{eq:P_a1}
\end{equation}
where $\theta_{i,S}$ is the critical inclination angle at which the observed flux is $S$; clearly, for $\theta_i < \theta_{i,S}$ the observed flux is $>S$.

\begin{figure*}
\centering{

\subfigure{\includegraphics[width=0.95\textwidth]{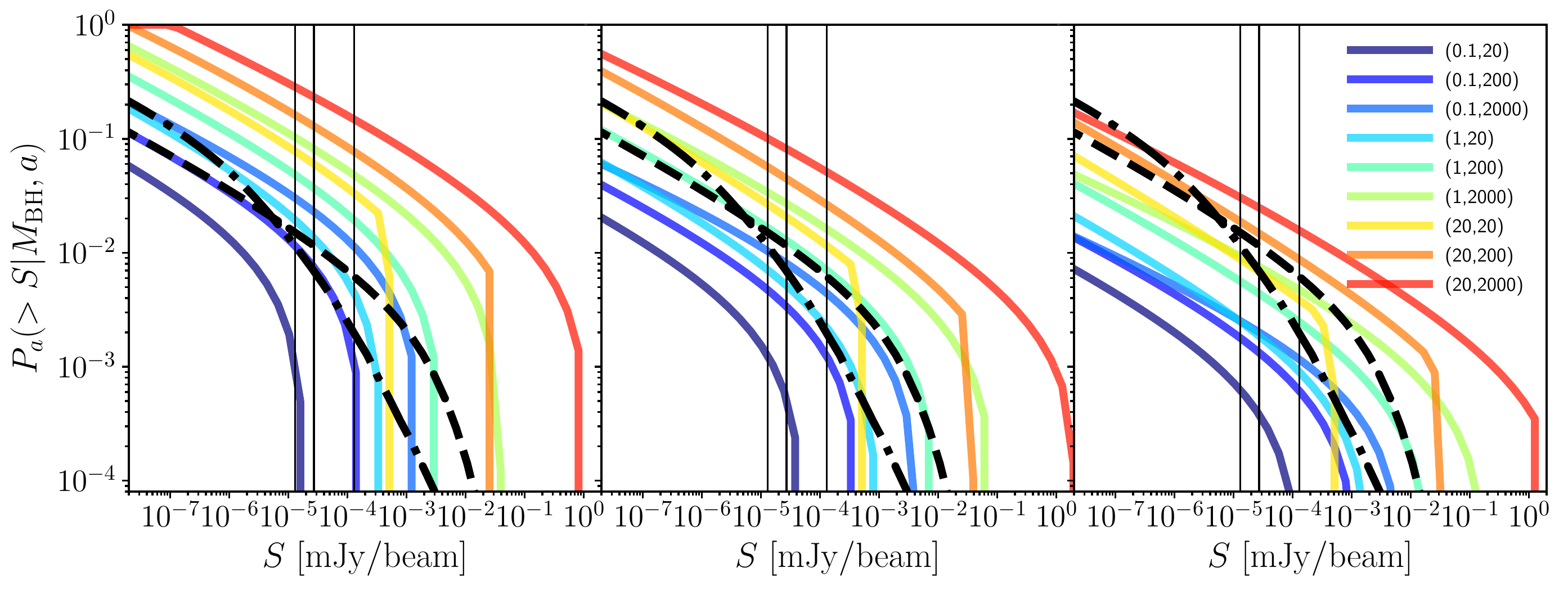}}
 \caption{The probability to observe a DCBH with flux $> S$ at 17.09 GHz. In each row, for left to right the panels correspond to $\Gamma=5.0$, 10.0 and 20.0 respectively. For each panel, the $B$ and $r_{\rm blob}$ of each curve are given in parentheses in the right panel. The dashed and dashed-dotted lines are probabilities after marginalizing the $B$, $\Gamma$ and $r_{\rm blob}$ for d1 distribution model and d2 distribution model respectively. In each panel vertical lines refer to sensitivities of SKA2-mid (left),  ngVLA (middle) and SKA1-mid (right) respectively.
}
\label{fig:P_S_BH1}
}
\end{figure*}

The probability distribution given by Eq. (\ref{eq:P_a1}) depends on jet properties that are unfortunately not known for DCBHs. In Fig. \ref{fig:P_S_BH1} we first plot $P_a(>S)$ given jet parameters: $B=(0.1, 1, 20)$ G, $\Gamma=(5,10, 20)$ and $r_{\rm blob}=(20,200,2000)$ $r_S$, to show how it changes with varying jet properties. We set $M_{\rm BH}=10^6~M_\odot$, $a=0.9$ and $z=10$, and observed frequency $\nu_{\rm obs}=17.06$ GHz.
For simplicity we fix the form of injected electrons distribution within the blob,  say $\gamma_{\rm min}=1$, $\gamma_{\rm break}=3\times10^2$ and $\gamma_{\rm max}=5\times10^3$, $s_1=1.0$ and $s_2=3.0$, only adjust its normalization according to the $P'_{\rm i}$ that is derived from $P_{\rm jet}$.  

In each panel of Fig. \ref{fig:P_S_BH1} we mark the SKA  sesitivity at
17.06 GHz and ngVLA sensitivity at 16 GHz. For all sensitivities we
assume 100 integration hours and, for SKA we have bandwidth $\Delta
\nu/\nu_c=0.3$,  and for ngVLA $\Delta \nu/\nu_c=0.5$. From this
figure, for SKA-mid and ngVLA the black hole is detectable in a large
parameter space, i.e., as long as $B\gtrsim 0.1$ G, $\Gamma \gtrsim 5$
and $r_{\rm blob}\gtrsim20$ $r_{\rm S}$. We also check that for
SKA-low (not shown in the figure) the the black hole is not detectable
for all above parameters. This is because when the black hole has
large flux, generally it has small inclination angle. As shown in
Fig. \ref{fig:Fobs_jet-GT09}, for  small inclination angle the low
frequency radiation is heavily absorbed due to the SSA effect. In
particular, for higher $\Gamma$, the probability to detect a high flux is higher; however in this case the energy is more concentrated on a narrow beam, therefore the probability to detect low-frequency flux is even smaller. The combined SKA-low and SKA-mid  measurements could be useful for revealing the jet properties.

If the probability distributions of the three jet parameters are known, then we can marginalize them to get a probability that only depends on black hole mass and spin,
\begin{align}
&P_a(>S|M_{\rm BH},a)=  \nonumber \\
& \int d Bd\Gamma d r_{\rm blob}     f(B)f(\Gamma)f( r_{\rm blob})  P_a(>S|M_{\rm BH},a,B,\Gamma,r_{\rm blob}).
\end{align}
Unfortunately, for DCBHs $f(B)$, $f(\Gamma)$ and $f(r_{\rm blob})$ are
only vaguely known; lacking a better argument,   we make some
assumptions. The assumed distributions do not have physical
motivations, and they are rather empirically inspired; however we hope
that they can provide at least a semi-quantitative estimate of the detectability. 
 
First,  the combination of the \citet{Ghisellini2010,Ghisellini2011,Ghisellini2013,Ghisellini2015,Paliya2017,Paliya2019,Paliya2020} observed samples shows that the $\log B$, $\Gamma$ and $\log r_{\rm blob}$ follow distributions similar to Gaussian shape, with  $\mean{\log B}=0.2, \sigma_{\log B}=0.2$, $\mean{\Gamma}=11, \sigma_{\Gamma}=2.5$, and $\mean{\log r_{\rm blob}}=1.9, \sigma_{\log r_{\rm blob}}=0.3$. Since there is no any prior knowledges about jets of DCBHs, we borrow these statistics directly. This is model d1. The results for $\nu_{\rm obs}=17.09$ GHz are added to the panel of Fig. \ref{fig:P_S_BH1} as a dashed black curve. 

Model d1 is based on properties of observed samples. Since in principle weaker jets are more difficult to detect, it is possible that  compared with the intrinsic distributions the observed jets are biased to stronger samples. On the other hand, power-law distribution $f(\Gamma)\propto \Gamma^{-2}$ for $\Gamma$ is found for some observed samples, e.g., \citet{Saikia2016}. So we also adopt another assumption that  $B$, $\Gamma$ and $r_{\rm blob}$ follow power-law distributions with index -2. This is model d2. The results for $\nu_{\rm obs}=17.09$ GHz are shown by dashed-dotted curves in Fig. \ref{fig:P_S_BH1}. 
 
We limit our parameters in the ranges: $0.1 \le B \le 20$ G, $2 \le \Gamma \le 30$  and $10 \le r_{\rm blob}\le 4000$ $r_{\rm S}$, roughly the boundaries of the above observed samples. At the lower limits the jet is already very weak for radio detection, so it is not necessary to investigate parameters even smaller than these lower limits. Beyond the upper limits, objects would be too rare in both d1 and d2 distributions. The results of the probabilities  are actually not sensitive to the upper limits.

\begin{figure}
\centering{
\subfigure{\includegraphics[width=0.5\textwidth]{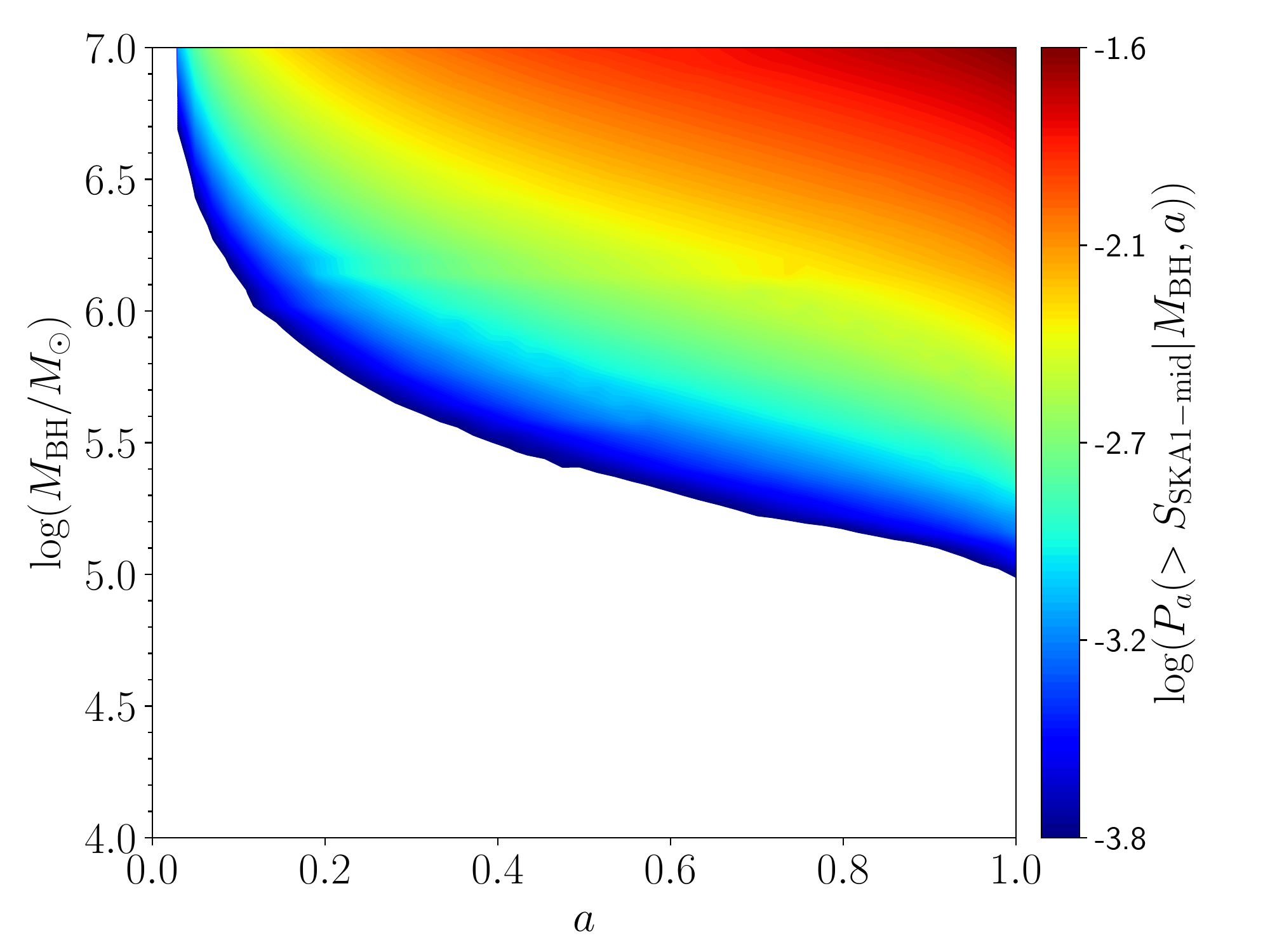}}
\subfigure{\includegraphics[width=0.5\textwidth]{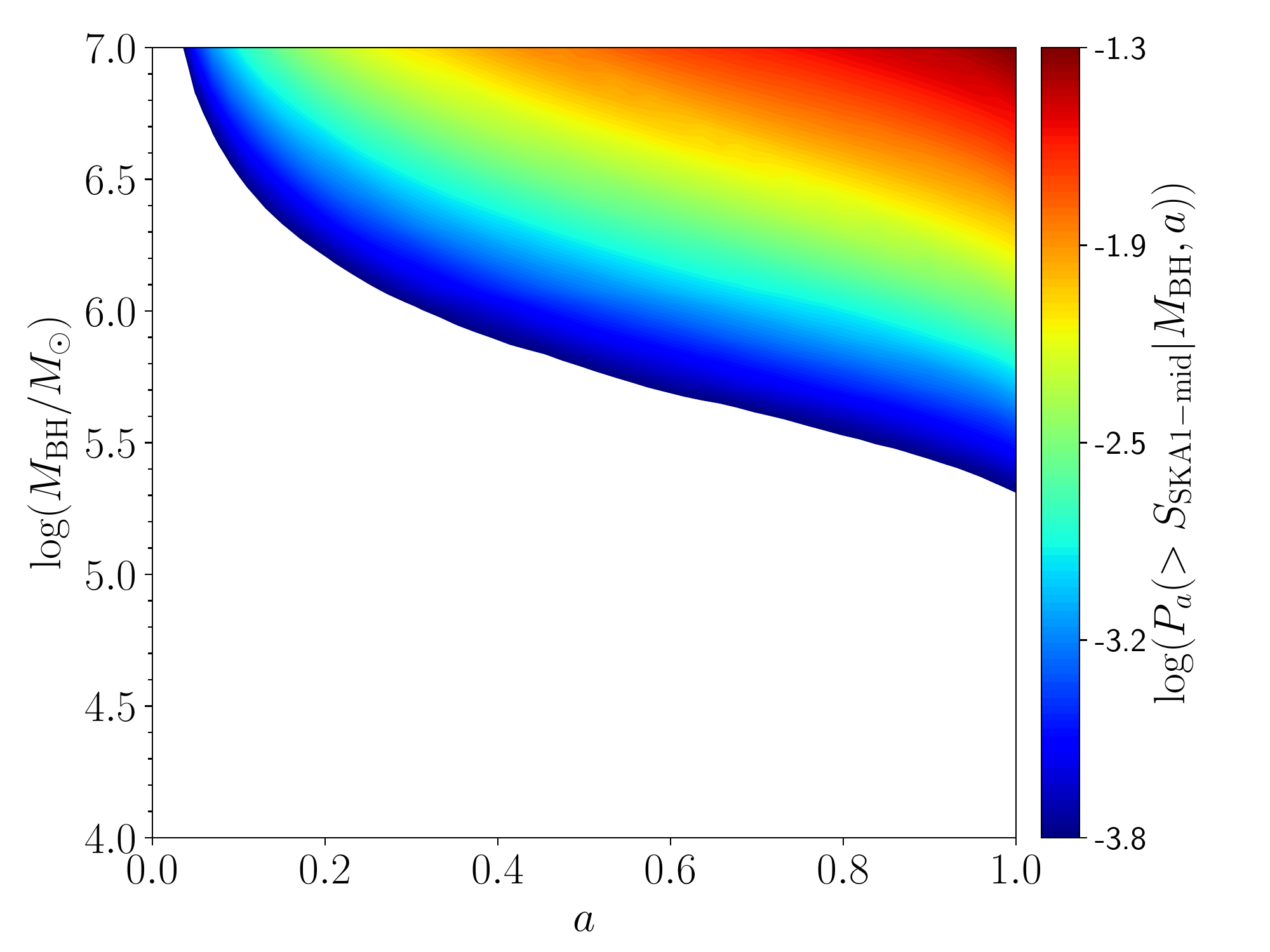}}
\caption{The probability of a BH with  flux above SKA1-mid sensitivity with 100 hours integration time (at 17.09 GHz with $\Delta \nu/\nu_c=0.3$), as a function of BH mass and spin. {\it Top}: For model d1. {\it Bottom}: For model d2.
}
\label{fig:P_S_BH2_grid}
}
\end{figure}

We further calculate the detectability by SKA1-mid at 17.09 GHz for varying the black hole mass and spins.
The results for  model d1 (top) and model d2 (bottom) are plotted in Fig. \ref{fig:P_S_BH2_grid}. We find that  black holes with
$M_{\rm BH}\gtrsim10^5~M_\odot$ or $\gtrsim3\times10^5~M_\odot$ are detectable in this case. Of course the results depend on the jet properties distributions and currently the distribution model d1 and d2 are pure assumptions.

We next consider marginalizing the spin distribution $f_{\rm spin}(a)$, 
\begin{align}
P(>S|M_{\rm BH}) &= \int da f_{\rm spin}(a) P_a(>S|M_{\rm BH},a) \nonumber \\
  &= \int da f_{\rm spin}(a) \frac{[1-{\rm cos}\theta_i(S,M_{\rm BH},a)]}{2},
\end{align}
to get the possibility just depends on black hole mass, where $f_{\rm spin}(a)$ is the normalized spin distribution.

Observations support the idea that most black holes have high spins, see \citet{Reynolds2019}. However, as high spin black holes are more easily selected, this evidence might be affected by a bias. On the other hand, the intrinsic spin distribution is still poorly known. \citet{Tiwari2018} suggested three possible spin distributions:
\begin{equation} f_{\rm spin}(a)=
\begin{cases}
2(1-a) ~~~{\rm favoring~low~spin }\\
1 ~~~ ~~~~~~~~~~~{\rm flat~spin~distribution  } \\
2a~~~~~~~~~~~~{\rm favoring~high~spin }
\end{cases}
\end{equation} 
 
\begin{figure}
\centering{
\subfigure{\includegraphics[width=0.5\textwidth]{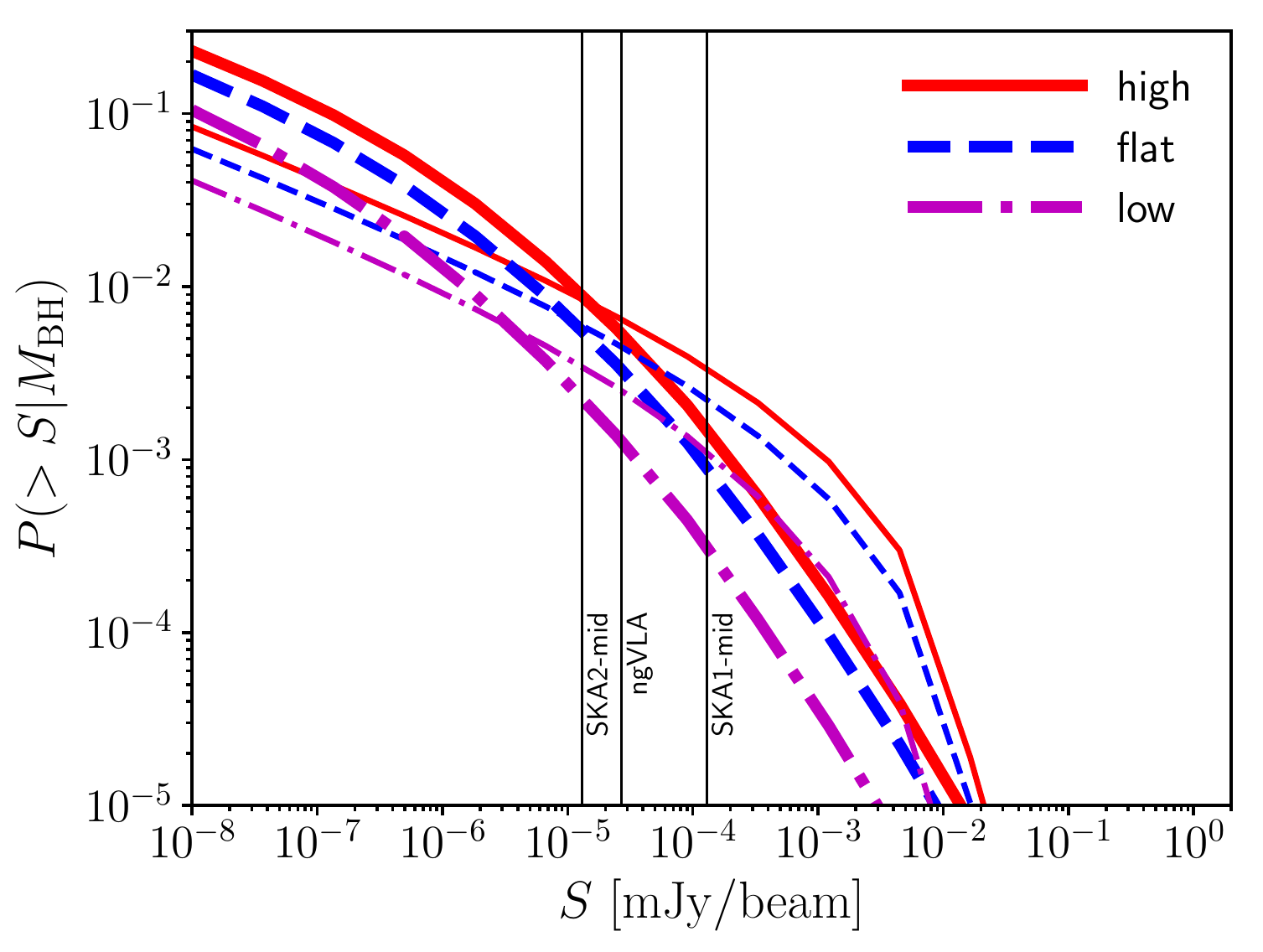}}
\subfigure{\includegraphics[width=0.5\textwidth]{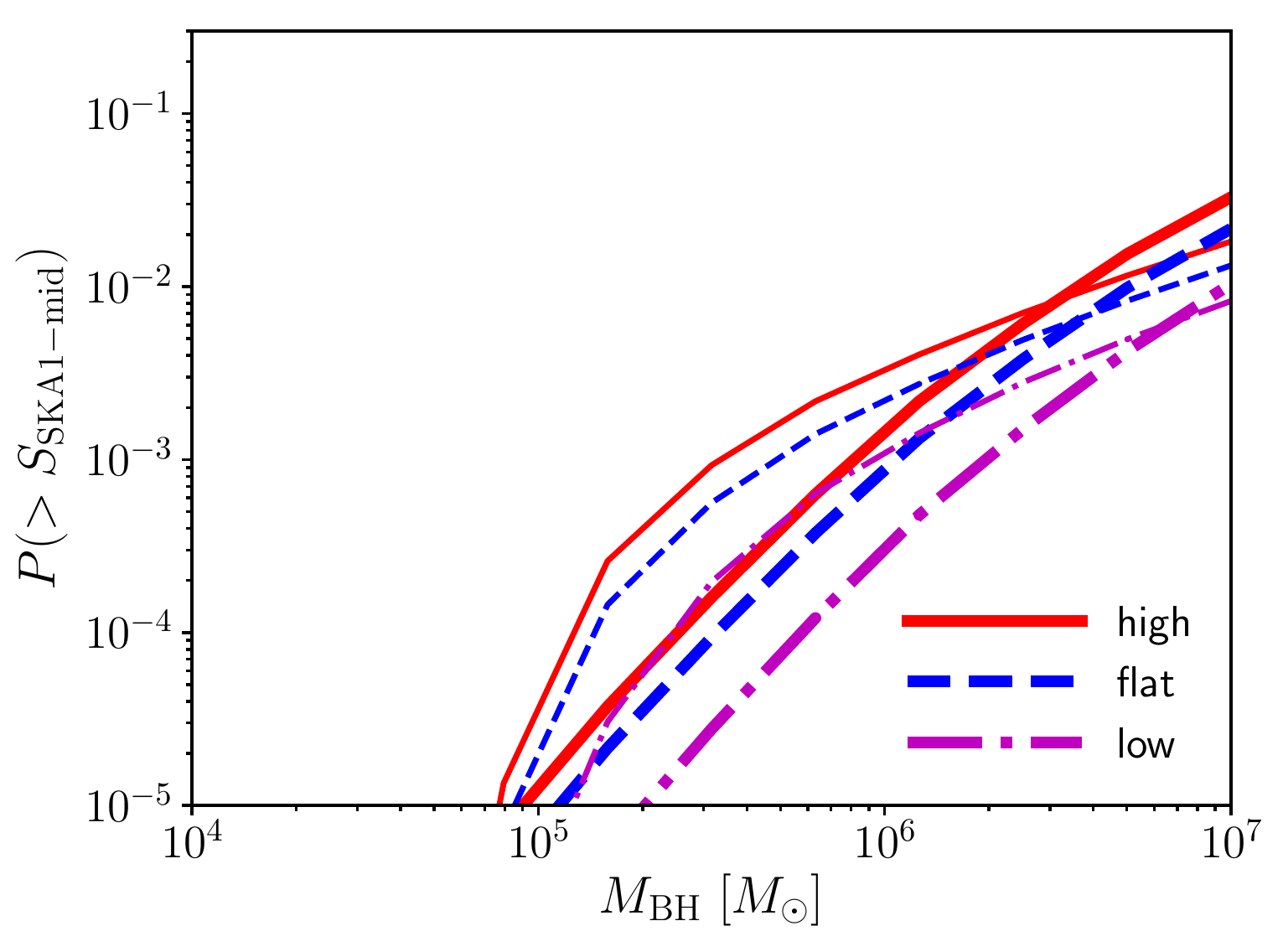}}
\caption{{\it Top:} The detection probability for a $10^6~M_\odot$ black hole with observed flux $> S$ for 17.09 GHz, given the spin distribution models indicated in the panel..
Thin lines are for $(B,\Gamma,r_{\rm blob})$ distribution model d1 and thick lines are for model d2.
 Vertical lines show the sensitivity for SKA and ngVLA, as in Fig. \ref{fig:P_S_BH1}.
{\it Bottom:} The detection probability by SKA1-mid at 17.09 GHz, as a function of black hole mass.
}
\label{fig:P_S_BH2}
}
\end{figure}

For the above spin distribution,  the observed flux probability at 17.09 GHz from a $10^6~M_\odot$ black hole is plotted in the top panel of Fig. \ref{fig:P_S_BH2}, assuming jet properties distribution as model d1 and model d2. In the bottom panel of Fig. \ref{fig:P_S_BH2} we plot the probability for a DCBH to be detected by SKA1-mid at 17.09 GHz in 100 integration hours, as a function of black hole mass. We find that, considering the spin distribution, roughly fraction $\sim10^{-3}$ of DCBHs with mass $10^6~M_\odot$ is detectable by SKA1-mid, if they all launch jets.  

 Finally, for DCBHs, if the mass function is known, then the number density above a flux $S$ is
\begin{equation}
n_{\rm BH}(>S)=\int d\log M_{\rm BH} \frac{dn}{d\log M_{\rm BH}}P(>S|M_{\rm BH}),
\end{equation}
Until now there are two theoretical investigations about the DCBH mass function in literatures, see \citet{Lodato2007,Ferrara2014}.
Here we follow the results in \citet{Ferrara2014} who gives that the mass function is a bimodal Gaussian form\footnote{In \citet{Ferrara2014} the mass function evolves as the accretion proceeds, and it depends on whether stars can form in minihalos or not. This formula is for the mass function at the end of accretion, in the model where stars can form in minihalos.}, i.e. $\propto \exp[-(\log M_{\rm BH} -x_1)^2/2 \sigma_1^2]+\exp[-(\log M_{\rm BH} -x_2)^2/2 \sigma_2^2]$, where $x_1=5.1, \sigma_1=0.3, x_2=5.9, \sigma_2=0.1$.

Regarding the normalizations for the total DCBH number density, from \citet{Dijkstra2014}, if the critical radiation $J_{21}=30$, then at $z\sim10$ the total number density of DCBHs is $\sim 2.5\times10^{-3}$ Mpc$^{-3}$. This is close to \citet{Habouzit2016} where for critical radiation $J_{21}=30$ the DCBH number density is $\sim1.0\times10^{-3}$ Mpc$^{-3}$. If however the critical radiation is 300, then  the total number density is $\sim1.7\times10^{-7}$ Mpc$^{-3}$ in  \citet{Dijkstra2014}.

DCBHs could be even rarer than found above, since there are feedback effects impact their formation. For example, for a realistic spectrum of external radiation from normal star-forming galaxies, the critical $J_{21}$ could be as high as $\sim10^3-10^4$ \citep{Sugimura2014,Latif2015}. During the formation of DCBHs the negative effects, such as the X-ray ionization/heating \citep{Inayoshi2015,Regan2016} and the tidal disruption \citep{Chon2016}, would significantly reduce the formation rate. 

However, there are also positive effects, such as dynamical heating
\citep{Wise2019}, H$^-$ detachment by Ly$\alpha$ \citep{Johnson2017},
and cooling suppression by radiation from other previously-formed
DCBHs \citep{Yue2017}, that can enhance the DCBH formation. Moreover,
if indeed DCBHs are seeds of SMBHs, their number density should be at
least  $\gtrsim$ of the number density of SMBHs. From some observations in recent years, at $z\sim6$ the observed QSOs number density  is $\sim10^{-7}$ Mpc$^{-3}$ (e.g. \citealt{Willott2010,Kashikawa2015,Faisst2021}). Theoretical models based on these observations predict that the QSOs number density reaches $\sim10^{-3}-10^{-4}$ Mpc$^{-3}$ down to $-15$ absolute UV magnitude \citep{Manti2017,Kulkarni2019}.

\begin{figure}
\centering{
\subfigure{\includegraphics[width=0.5\textwidth]{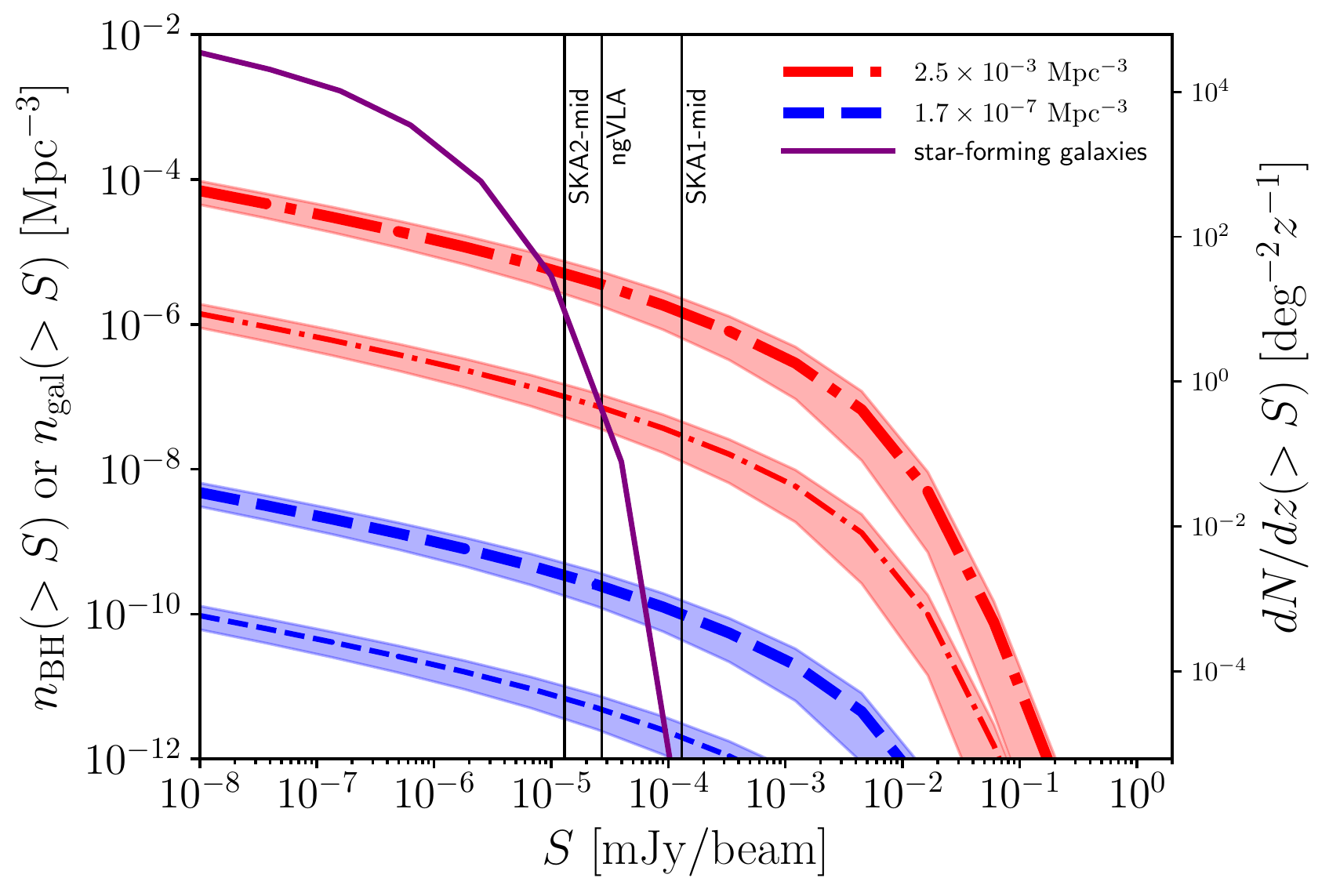}}
\subfigure{\includegraphics[width=0.5\textwidth]{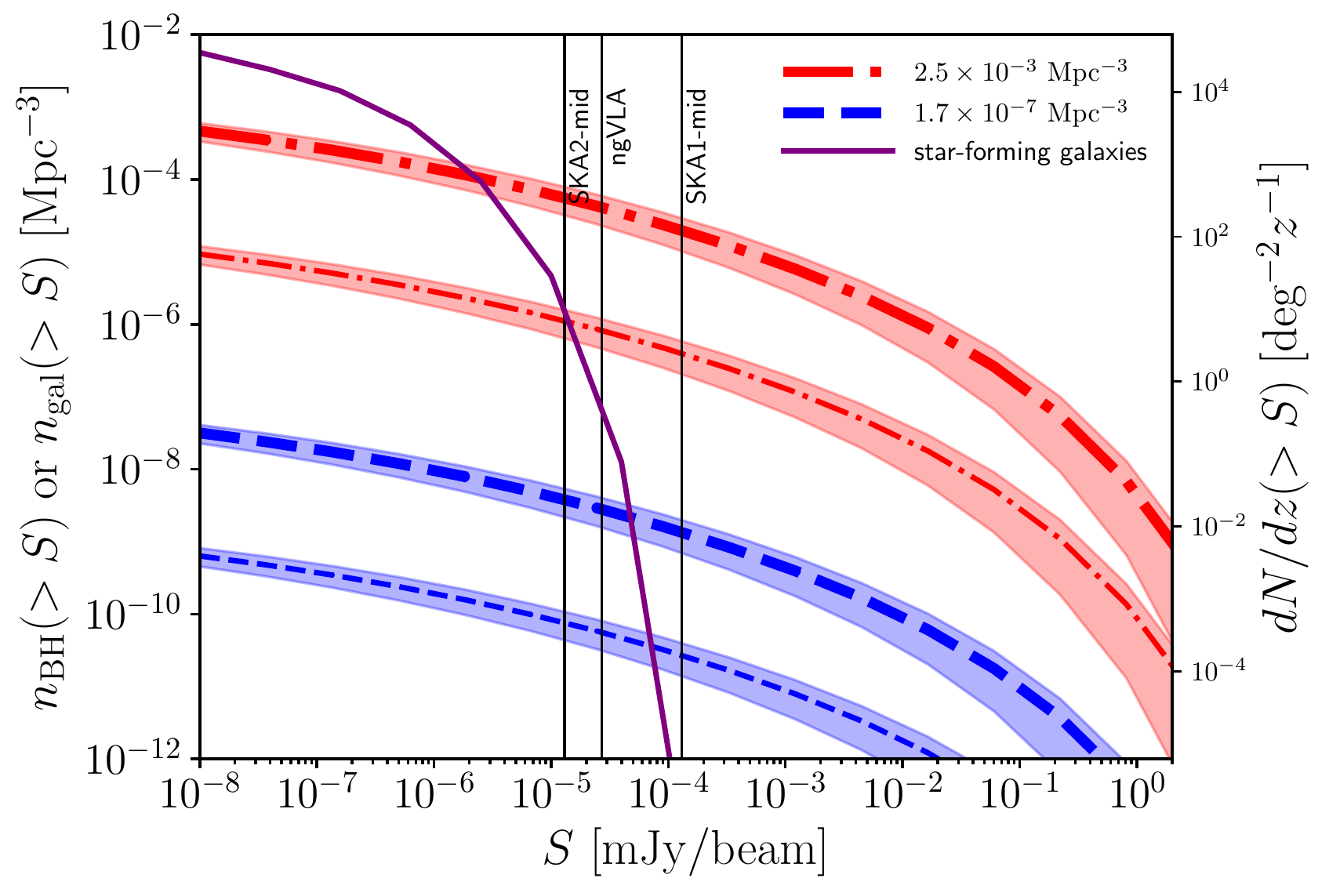}}
\caption{
The  number density of DCBHs with flux $>S$ at $z=10$, for model d1 (top) and model d2 (bottom) respectively, for total DCBH number density $2.5\times10^{-3}~$Mpc$^{-3}$ (dashed-dotted) and $1.7\times10^{-7}~$Mpc$^{-3}$ (dashed) respectively. We assume the black hole mass follows the \citet{Ferrara2014} distribution. Thick curves are for a duty cycle $100\%$ while thin curves are for duty cycle $2\%$. 
Curves are for flat spin distribution, around each line, the filled regions mark the ranges from favoring low-spin to favoring high-spin distribution. We also plot the  number density of of star-forming galaxies with synchrotron flux $>S$ at $z=6$ (solid).  Right y-axis is the surface number density. 
}
\label{fig:n_BH_Gal}
}
\end{figure}

To predict the DCBH number count, one should specify a total number density as normalization. Currently this normalization is quite uncertain, as explained above. In Fig. \ref{fig:n_BH_Gal}  we plot the  $n_{\rm BH}(>S)$ at $z=10$, for total DCBH number density normalization of $2.5\times10^{-3}$ Mpc$^{-3}$ and $1.7\times10^{-7}$ Mpc$^{-3}$  respectively. We convert the number density into surface number density in the right $y$-axis. For other number density of DCBHs, the curves should be re-normalized accordingly.

Generally for low-luminosity radio galaxies (classified as ``FR-I'') the jet lifetime could be as long as the Hubble time because the central black hole is fueled at a low rate \citep{Blandford2019ARAA}. while for high-luminosity FR-IIs the duty cycle is $\simeq 2\%$, and each jet episode lasts only $\simeq 15$ million years  \citep{Bird2008}. Motivated by this, in Fig. \ref{fig:n_BH_Gal} we also plot $n_{\rm BH}(>S)$ assuming a $2\%$ duty cycle by thin lines. In this case the number counts are considerably reduced. 

We remind that in Fig. \ref{fig:n_BH_Gal} it is assumed that all DCBHs are jetted. If however, only a fraction of them can launch the jet, then the number count or luminosity function should be re-normalized according to the jetted fraction.

We also estimate the radio flux from star-forming galaxies, to check if it would confuse the DCBH observation. For a star-forming galaxy,  the synchrotron luminosity is related to the star formation rate (SFR), because the electrons are accelerated by shocks from supernova explosions whose rate is proportional to the SFR.  A relation is given:  (see \citealt{Bonato2017} and references therein)
\begin{align}
\frac{ \bar{L}_{\rm sync}(\nu)}{{\rm erg~s^{-1}Hz^{-1}}}&=1.9\times10^{28} \left(\frac{\rm SFR}{M_\odot \rm yr^{-1}} \right) \left(\frac{\nu}{\rm GHz}\right)^{-0.85}
\left[1+\left(\frac{\nu}{\rm 20GHz}\right)^{0.5}\right]^{-1}, 
\end{align}
and after the low luminosity correction,
\begin{equation}
L_{\rm syn}(\nu)=\frac{ L_{\rm syn}^*    } {   (L_{\rm syn}^*/\bar{L}_{\rm syn} )^3+(L_{\rm syn}^*/\bar{L}_{\rm syn} ) },
\end{equation}
where $L_{\rm syn}^*=0.886\bar{L}_{\rm syn}({\rm SFR}=1~M_\odot{\rm yr}^{-1})$.

On the other hand, the SFR is also proportional to the UV luminosity \citep{Yue2019}, 
\begin{equation}
\frac{\rm SFR}{M_\odot {\rm yr}^{-1}}\approx0.7\times10^{-28} \frac{ L_{\rm UV} }{ \rm erg~s^{-1}Hz^{-1}   },
\end{equation}
we therefore can construct the radio luminosity function from the UV luminosity function via
\begin{equation}
\frac{dn}{d\log L_{\rm syn}}=\frac{dn}{d\log L_{\rm UV}}\frac{d\log L_{\rm UV}}{d\log L_{\rm syn}}.
\end{equation}
The dust-corrected UV luminosity function $dn/d\log L_{\rm UV}$ is given by \citet{Bouwens2014,Bouwens2015}.

In Fig. \ref{fig:n_BH_Gal} we also plot the number density of the star-forming galaxies at $z=6$. We conclude that, although DCBH might be rarer than star-forming galaxies, they still dominate the brightest-end of the flux distribution. Restricting to sources with continuum flux $\gtrsim10^{-4}$ mJy would prevent confusion between DCBH and star-forming galaxies.

\section{Summary and discussion}\label{sec:summary}

We have explored the possibility to detect the radio signal from DCBHs by upcoming radio telescopes, such as the SKA and ngVLA. 
We just consider the mechanisms that produce radio signal  at the accretion stage after DCBH formation. DCBHs may also produce radio signal during the collapse stage. However this stage is shorter therefore the detection probability is smaller.
Our model supposes that DCBHs can launch powerful jets similar to those observed in the radio-loud AGNs. We applied the jet properties of the observed blazars to the jetted high-$z$ DCBHs, and used the publicly available GT09 model to predict the radio signal from the jet. We found that:

\begin{itemize}
    \item If the jet power $P_{\rm jet}\gtrsim10^{42-43}$ erg s$^{-1}$, the continuum radio flux can be detected by SKA and ngVLA with 100 integration hours, depending on the inclination angle with respect to the line-of-sight. 
    \item However, as the jet power depends on both black hole mass and spin, if the black hole mass $M_{\rm BH}\lesssim1\times10^5~M_\odot$, even maximally spinning DCBH can be hardly detected by SKA1. 
    \item By  considering the spin distribution, about $10^{-3}$ of DCBHs with $M_{\rm BH}=10^6~M_\odot$, if they are all jetted, are detectable.
    \item Moreover, if the blob at the jet head is dense ($r_{\rm blob}\sim 50~r_S$), and has large bulk motion velocity ($\Gamma\sim10$), then it may only feebly emit in the SKA-low band, because of the SSA and radiation blueshift. These events are more suitable for study with SKA-mid. However, by combing the observations of SKA-low and SKA-mid, this may also provide a potential to distinguish a jetted DCBH from a normal star-forming galaxies.
    \item  If all DCBHs are jetted, for the most optimistic case $\sim 100$ deg$^{-2}z^{-1}$ at $z=10$ would be detected by SKA1-mid with 100 hour integration time.
\end{itemize}

We have predicted that some DCBHs can be detected by upcoming radio telescopes if their jets have small inclination angles. Radio flux from other unresolved DCBHs forms a diffuse radiation field that is part of the cosmic radio background (CRB). This might be connected with previous claims of an ``excess'' on the CRB (e.g. \citealt{Seiffert2011}), whose origin has been related with undetected black holes \citep{Ewall-Wice2018}. If confirmed, future CRB observations will 
provide an alternative tool for study DCBHs.

Our study inevitably contains some uncertainties. We list them
  below and discuss their impact briefly.

\begin{enumerate}

\item Jetted fraction:

Although DCBH may launch a jet, it is not clear how many of DCBHs can
actually do this. Our predicted probabilities are for jetted DCBHs. 
So if some DCBHs do not show jets, or their jets are weaker than our lower boundaries  (i.e., $B < 0.1$ G, $\Gamma < 2$ and $r_{\rm blob}< 10$ $r_{\rm S}$), then the denominator  is higher and our probabilities should be re-normalized. However the probability-flux shape at the high-flux end is not influenced by normalization.

\item Jet properties:

To predict the observed radio flux from individual jetted DCBH, we have used the simple empirical $P_{\rm jet}-L_{1.4}$ relation, with reasonable assumed parameters, for example the jet power is $\lesssim$ the Eddington luminosity, and a typical Lorentz factor for the jet head is $\sim10$, see Eq. (\ref{eq:P_jet-L_radio}) and Fig. \ref{fig:S_obs}.  This prediction is not quite sensitive to the details of the jet mechanism. 

We then predicted the observed flux expanding the GT09 jet model. The
results are consistent with the above simple empirical relation, see
Fig. \ref{fig:Fobs_jet-GT09}.  We conclude that without using extreme
assumptions, some jetted DCBHs can be detected by SKA or ngVLA.

Further predictions on the detection probability suffer from
additional uncertainties. These predictions are based on jet property parameters borrowed from observed blazars, and the assumed distribution forms (Gaussian or power-law). Currently we cannot further improve such predictions. However, our estimations are at least useful as optimistic references.

\item Number density of the DCBHs and their mass function:
 
This is also an uncertain aspect, although its study is beyond the
scope of this paper. The number density of DCBHs predicted in
literature span $\sim$8 orders of magnitude.  The lower limit is close to
the number density of the observed high-$z$ SMBHs, which is $\sim10^{-9}$ Mpc$^{-3}$;
the upper limit is close to the number density of newly-formed
atomic-cooling halos which is $\sim0.01-0.1$ Mpc$^{-3}$, depending on the
critical radiation field, and how negative feedback mechanisms
work. In Fig. \ref{fig:n_BH_Gal} we show the number counts
corresponding to total number $\sim10^{-3}$ Mpc$^{-3}$ and
$\sim10^{-7}$ Mpc$^{-3}$. The former would be close to upper
limits. Regarding the mass function, although there is no
observational support, it is generally believed that the mass
distribution peaks at $10^5-10^6~M_\odot$. We have made our choice accordingly.

\end{enumerate}

In this work we attempt  to present some tentative investigations, we hope it can draw attention to observing these first black holes by SKA and ngVLA. 
Future theoretical investigations, for example the numerical simulations, would help to improve our predictions. 
However, the uncertainties could only be reduced  by observations themselves. For example, if the radio signal from DCBHs is much lower than our predictions, then it is likely that most DCBH cannot launch strong jet. This will force us to investigate the difference between DCBHs and low-$z$ black holes, for example on the magnetic field, the density of the envelope, or the spin.

 \section*{Acknowledgments} 
 
We thank Dr. Erlin Qiao and Dr. Xiaoxia Zhang for helpful discussions. BY acknowledges the support by the National Natural Science Foundation of China (NSFC) grant No. 11653003, the NSFC-CAS  joint fund for space scientific satellites No. U1738125, the  NSFC-ISF joint research program No. 11761141012. AF acknowledges support from the ERC Advanced Grant INTERSTELLAR H2020/740120. Any dissemination of results must indicate that it reflects only the author’s view and that the Commission is not responsible for any use that may be made of the information it contains. Partial support from the Carl Friedrich von Siemens-Forschungspreis der Alexander von Humboldt-Stiftung Research Award is gratefully acknowledged (AF).
 
\section*{Data availability} 
The data underlying this paper can be shared on reasonable request to the corresponding author.
 
\bibliographystyle{mn2e}
\bibliography{refe,refe2}

\appendix
\section{The synchrotron radiation calculation details}\label{sec:syn}

The synchrotron power per frequency emitted by a single particle with Lorentz factor $\gamma$ \citep{Strong2011,DiBernardo2013,DiBernardo2015,Ghisellini2013} is
\begin{equation}
G_{\rm syn}(\nu,\gamma)=\frac{\sqrt{3}e^3B}{m_e c^2}F(\nu/\nu_c),
\end{equation}
where $B$ is the magnetic field, $e$ is the electron charge and $m_e$ is the electron mass,  and $c$ is the speed of light.  
The function 
\begin{equation}
F(\nu/\nu_c)=\frac{\nu}{\nu_c}\int_{\nu/\nu_c}^\infty K_{5/3}(y)dy,
\end{equation}
with peak frequency
\begin{equation}
\nu_c=\frac{3e}{4\pi m_e c}B \gamma^2,
\end{equation}
$K_l$ is the modified Bessel function of order $l$ of the second kind. $F$ is called synchrotron function, we used the fittings given by  \citet{Fouka2013}.

Suppose the electron number density distribution is $dN_e/d\gamma$,  then the synchrotron emissivity is
\begin{equation}
j_{\rm syn}(\nu)=\frac{1}{4\pi}\int G_{\rm syn}(\nu,\gamma)\frac{dN_e}{d\gamma}(\gamma) d\gamma,
\end{equation}
where $dN_e/d\gamma$ is provided in the GT09 code. 

\section{the black hole spin}\label{sec:BH-spin}

As the total jet power depends on spin, it is important to check whether the host dark matter halo or the accretion disk contain enough angular momentum to support a high-spin central black hole but still allows a DCBH to form.

The angular momentum of a dark matter halo is \citep{Lodato2007}
\begin{equation}
J_h=\lambda_h \frac{GM_h^{5/2}}{|E_h|^{1/2}},
\end{equation}
where $\lambda_h$ is the dimensionless spin parameter of the halo, $E_h=(1/2)W_h$ is the total energy of the halo and $W_h$ is the
gravitational potential energy. For a NFW profile \citep{Lokas2001}, 
\begin{equation}
W_h(s)=-W_\infty\left[1-\frac{1}{(1+Cs)^2}-\frac{2{\rm ln}(1+Cs)}{1+Cs}\right],
\end{equation} 
where $s=r/r_{\rm vir}$ is the halo radius in units of virial radius $r_{\rm vir}$, $C$ is the concentration parameter, 
\begin{equation}
W_\infty=\frac{cg^2(C)GM_h^2}{2r_v}
\end{equation}
is the potential energy when $s\rightarrow\infty$ and 
\begin{equation}
g(C)=\frac{1}{{\rm ln}(1+C)-C/(1+C)}.
\end{equation}
For reference, a dark matter halo with $M_h=10^8~M_\odot$ at $z=10$ has $J_h\approx 2\times10^{66}({\lambda_h}/{0.001})~{\rm erg\, s}$.

In \citet{Lodato2006,Lodato2007}, they gave the disk surface density profile $\Sigma(R)\propto R^{-1}$, both the disk size and the surface density profile normalization are derived by let the mass equal fraction $m_d$ of the halo and angular momentum equal fraction $j_d$ of the dark matter halo angular momentum. 
We determine an inner radius of the accretion disk, letting 
\begin{equation}
M_{\rm BH}=\int_0^{R_{\rm in}} \Sigma(R)2\pi R dR,
\end{equation}
therefore the maximum angular momentum available for BH accretion is
\begin{equation}
J_{\rm acc}\sim\int_0^{R_{\rm in}} \Sigma(R) 2\pi R R^2\omega_d dR,
\end{equation}
where $\omega_d$ is the angular velocity of the disk derived from 
\begin{equation}
j_d J_h=\int_0^{R_d} \Sigma(R) 2\pi R R^2\omega_d dR,
\end{equation}
where $R_d$ is the size of the accretion disk.

\begin{figure}
\centering{
\subfigure{\includegraphics[width=0.45\textwidth]{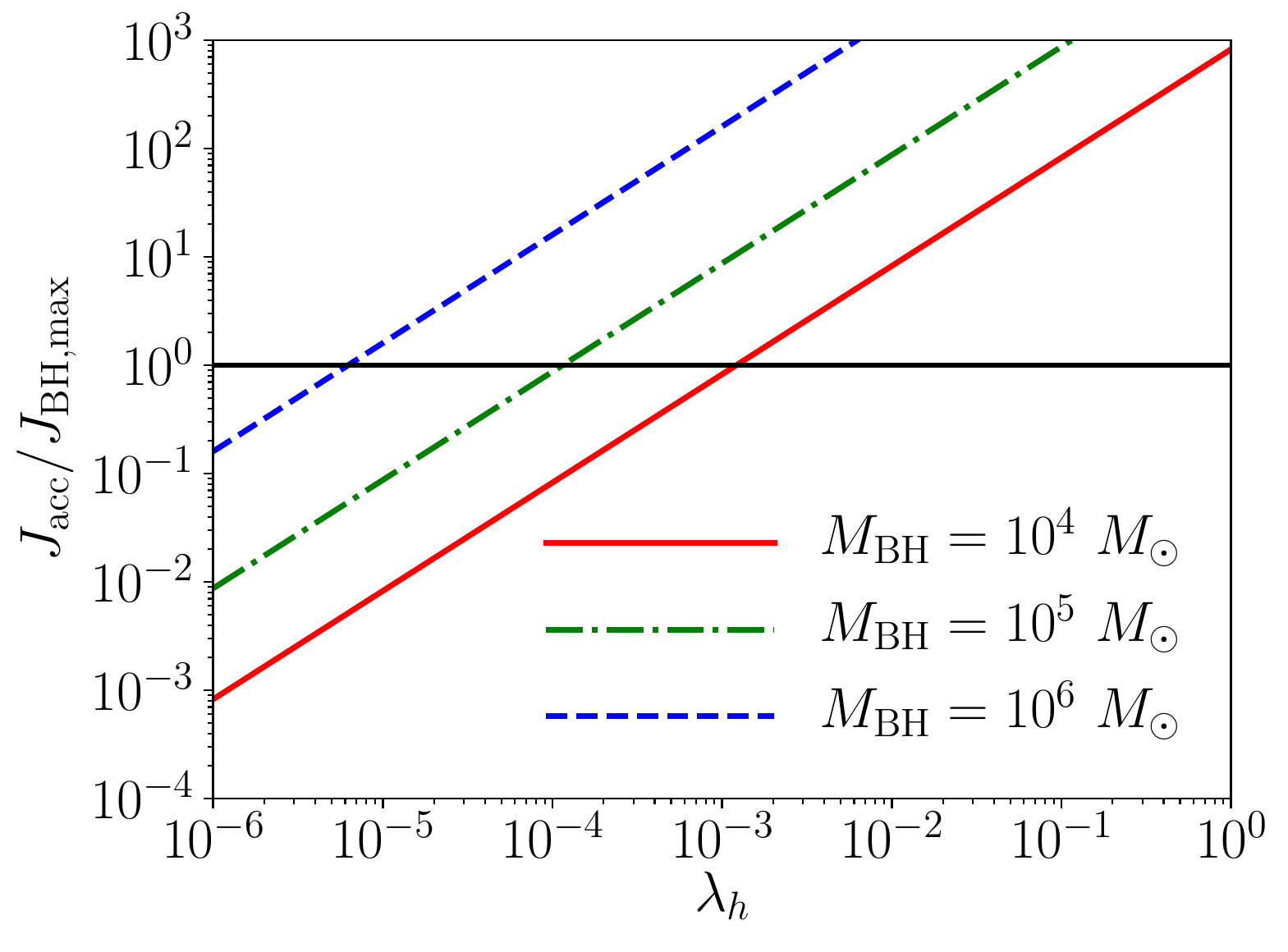}}
\caption{The $J_{\rm acc}/J_{\rm BH,max}$  as a function of $\lambda_h$ for different black hole mass.   
To guide the eye we plot a horizontal line corresponding to $J_{\rm acc}=J_{\rm BH,max}$.
 }
\label{fig:spin_acc}
}
\end{figure}

In Fig. \ref{fig:spin_acc} we plot the $J_{\rm acc}/J_{\rm BH,max}=J_{\rm acc}/(GM^2_{\rm BH}/c^2)$ as a function of $\lambda_h$, given the halo mass $M_h=10^8~M_\odot$, $z=10$, $m_d=j_d=0.05$. We find that, as long as the $\lambda_h\gtrsim10^{-3}-10^{-5}$, the accreted angular momentum is more than enough to support the maximum black hole spin. Since the spin parameter distribution of dark matter halos follows a ln-normal distribution, with central value ln$(0.05)$ and standard deviation 0.5, most dark matter halos have $\lambda_h\gtrsim 10^{-3}-10^{-5}$. A disk with high angular momentum is stable and supported by rotation, leading to a reduced central DCBH mass.  \citet{Lodato2006,Lodato2007} also give the relation between $\lambda_h$ and the central DCBH mass. We check that, $\lambda_h\sim10^{-3}-10^{-5}$ is small enough so that the formation of a massive central DCBH is possible.

Finally, theoretical calculations predict that if the BH spins up by acquiring the angular momentum of the innermost disk boundary throughout the accretion phase, for a thin disk,  the spin increases from 0 to $\approx 1$ in a short time scale,  roughly $\epsilon\times t_{\rm Edd}$, where $t_{\rm Edd}=45$ Myr is the Eddington time; generally the radiative efficiency $\epsilon\sim0.1$. Within this time scale the black hole doubles its mass (e.g., \citealt{Li2012,ZhangXiaoxia2019}).

\section{Radio signal of Radio-quiet DCBHs}\label{sec:radio-quiet}

Generally, radio-loud AGNs have higher spin than radio-quiet ones \citep{Wilson1995,Sikora2007}, and they are usually hosted by elliptical galaxies; radio-quiet AGNs are instead mostly found in spiral galaxies. However, even in radio-quiet AGNs there are mechanisms that produce some synchrotron radiation, such as, e.g. weaker jets, relativistic electrons in the corona, or winds/outflows (see the review e.g. \citealt{Panessa2019}). Shocks can also accelerate relativistic electrons and this process represents an additional possibility  for synchrotron \citep{Ishibashi2011}. Here, we do not model the complex radiative processes occurring in the vicinity of the black hole. Instead, we just assume some radio loudness parameter $R$, and, unlike the jet, now the synchrotron radiation is isotropic and we do not need to consider the beaming effect.  

In \citet{Yue2013b} we calculated the DCBH SED analytically. In the radio band the radiation is mainly free-free emission. Free-free emission has flat spectrum when the energy of photons is much smaller than the kinetic energy of the thermal electrons. At the low frequency, free-free absorption works, as described in Sec. \ref{sec:envelope}.

For Compton-thick DCBH, free-free absorption becomes important even at frequency as high as $\sim10^2-10^3$ GHz. As a result the  radio signal from such a DCBH is difficult to detect. We add the free-free absorption to \citet{Yue2013b}, assuming that the  density profile of the medium surrounding the central black hole is Eq. (\ref{eq:n_a}).

\begin{figure}
\centering{
\subfigure{\includegraphics[width=0.5\textwidth]{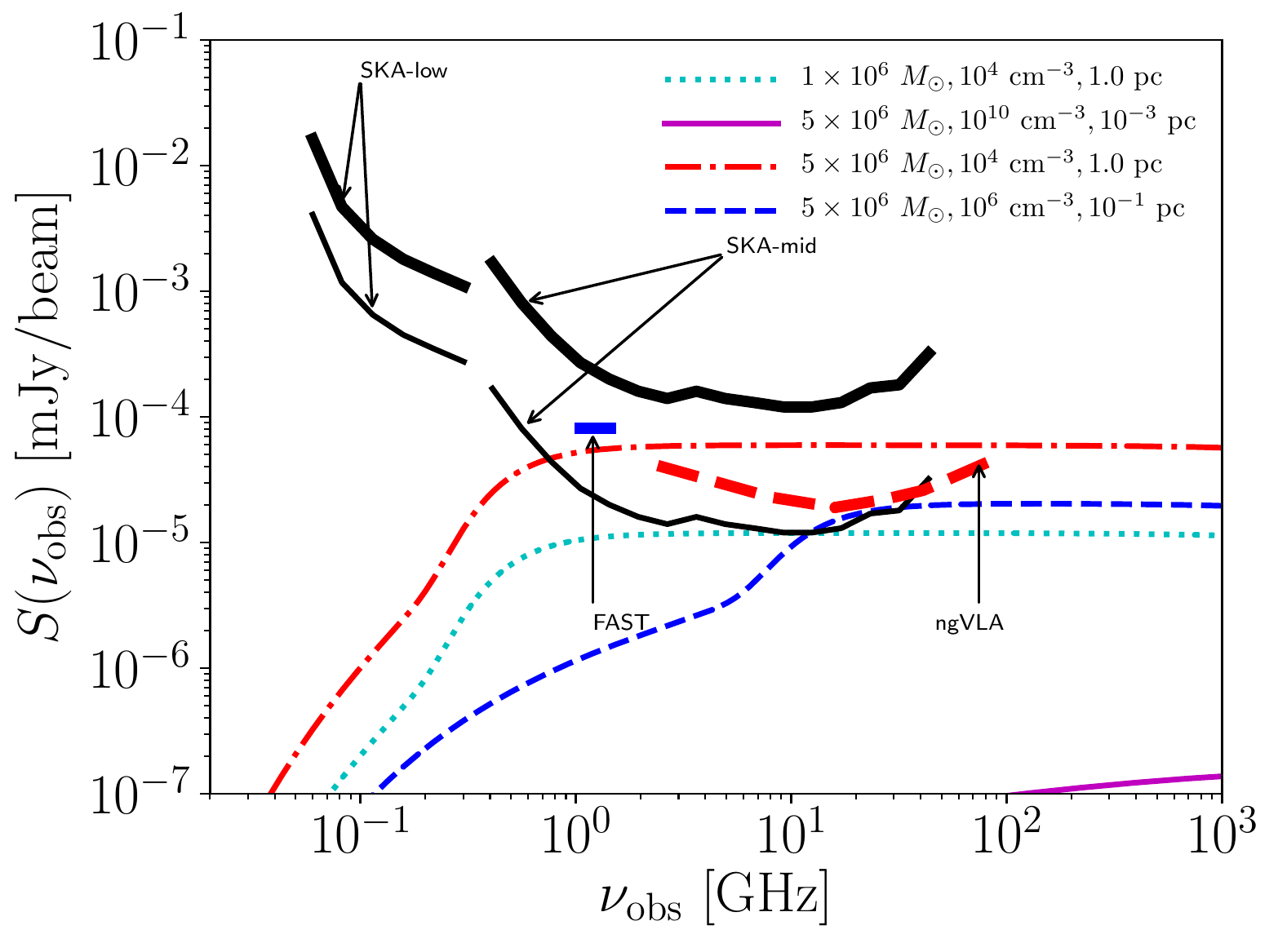}}
\caption{The flux from a $z=10$ DCBH with mass and $N_{\rm H}$ given in the legend, we assume the total luminosity is the Eddington luminosity. We show the sensitivities of SKA1(2)-low, SKA1(2)-mid and ngVLA as well, for integration time 100 hours.
}
\label{fig:Flux_DCBH}
}
\end{figure}

In Fig. \ref{fig:Flux_DCBH} we plot the flux of DCBH at $z=10$ with various black hole mass and envelope profile  (we assume $N_{\rm H}\sim N_{\rm e}$). We find that, without the synchrotron radiation, the radio-quiet DCBH is not able to detect by SKA-low and SKA1-mid. However, by SKA2-mid or ngVLA, it is marginally detectable if $M_{\rm BH}\gtrsim 5\times10^6~M_\odot$ and $N_{\rm H}\lesssim10^{23}$ cm$^{-2}$.

\begin{figure}
\centering{
\subfigure{\includegraphics[width=0.5\textwidth]{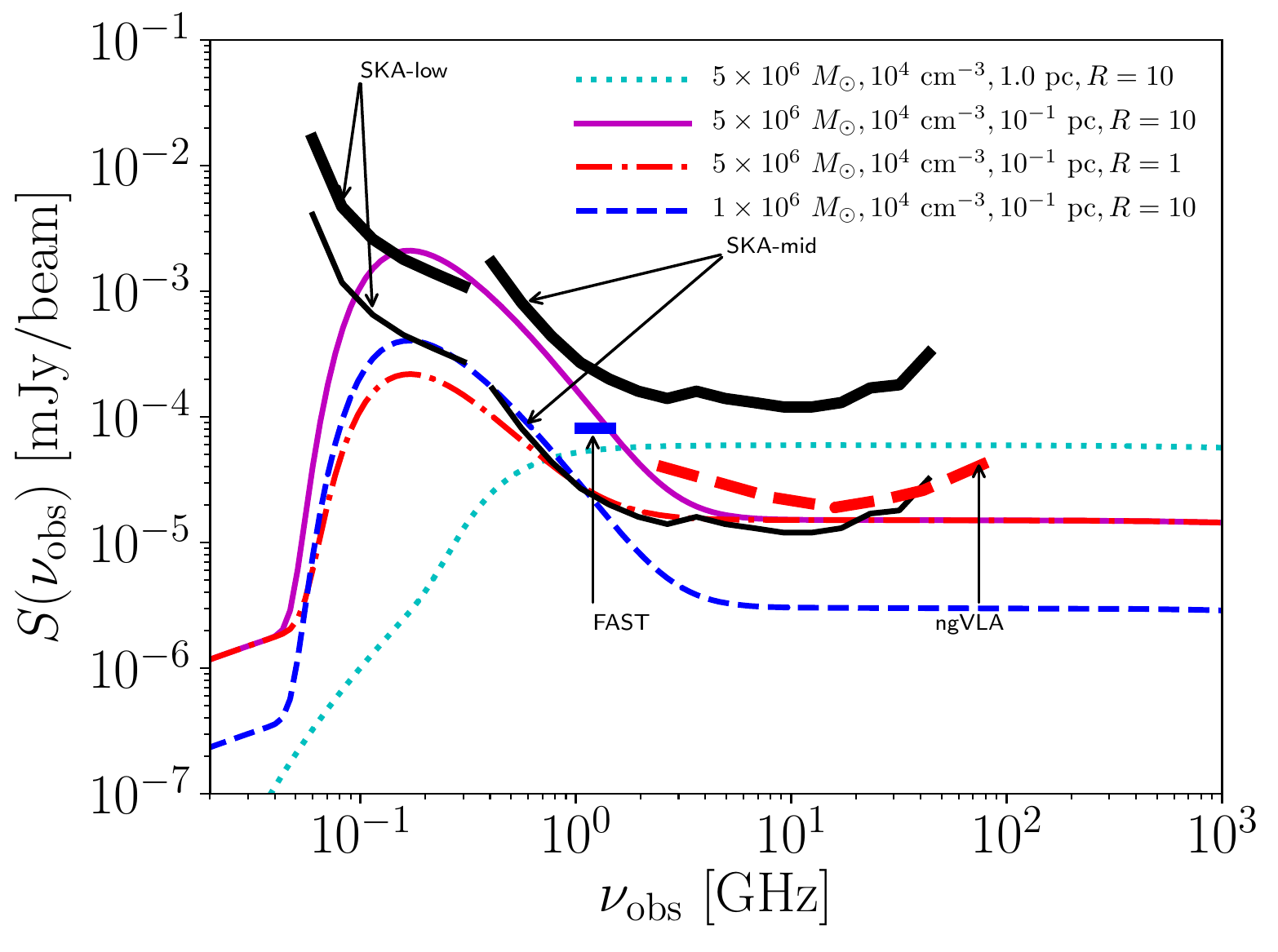}}
\caption{Similar to Fig. \ref{fig:Flux_DCBH}, however here we assume the radio-quiet DCBH also produce some synchrotron radiation. 
}
\label{fig:Flux_DCBH_syn}
}
\end{figure}

We then assume that such radio-quiet DCBH may also  produce some synchrotron by assuming a radio loudness parameter $R$. 
Here $R$ is the ratio between the flux observed at 5 GHz and at 4400 \AA. 
The results are shown in Fig. \ref{fig:Flux_DCBH_syn}. For a black hole with mass $\sim5\times10^6~M_\odot$ and $R=10$, the signal is marginally detectable by SKA-low.
Note that by the definition of the terminology ``radio-quiet'', $R=10$ is the maximum radio loudness parameter \citep{Kellermann1989}. So the latter is considered as an optimistic extreme. Again, detecting the synchrotron emission by SKA-low typically requires $N_{\rm H}\lesssim10^{22}$ cm$^{-2}$. 
 
\section{Jet break out and highly-relativistic blob formation}\label{sec:jet-prop}

Generally, the DCBH accretion disk is surrounded by an envelope that could be Compton-thick with column number densities up to $\sim10^{25}$ cm$^{-2}$ \citep{Yue13a}. Before investigating the DCBH radio detectability, it is necessary to study the jet propagation inside this envelope and its eventual break-out. 
We follow the theory of jet propagation given in \citet{Bromberg2011} (see also  \citealt{Matzner2003,Matsumoto2015}).

{If the jet energy loss is negligible,} the speed of the jet head (in units of $c$) is
\begin{equation}
\beta_h=\frac{\beta_j}{1+\tilde{L}^{-1/2}},
\end{equation}
where $\beta_j$ is the jet speed, and 
\begin{equation}
\tilde{L}\approx\frac{P_{\rm jet}}{\Sigma_h \rho_a c^3},
\end{equation}
here $\Sigma_h$ is the cross-section of the jet head; $\rho_a = m_{\rm H} n_a$ is the envelope mass density at the jet head, with $m_{\rm H}$ being the hydrogen mass (for simplicity we assume a pure H gas), for which is given by Eq. (\ref{eq:n_a}).

A jet could be either ``collimate'' or ``un-collimated'' (conical), depending on the competition between the jet pressure and the cocoon pressure. If the jet is un-collimated and the opening angle is $\theta_0$, then 
\begin{equation}
\Sigma_h=r_h^2\pi\theta_0^2,
\end{equation}
where the propagation distance of the jet head 
\begin{equation}
r_h(t)=\int_0^t c\beta_h dt.
\end{equation}

During jet propagation, a cocoon is produced, and the jet loses energy due to the cocoon expansion. The energy going into the cocoon is
\begin{equation}
E_c=\eta P_{\rm jet}\int_0^t(1-\beta_h)dt,
\end{equation}
where $\eta \lesssim 1$ is an efficiency.
One writes the volume, pressure and expansion velocity of the cocoon as 
\begin{equation}
V_c=\pi c^3\int \beta_h dt \left(\int\beta_cdt\right)^2,
\end{equation}
\begin{equation}
P_c=\frac{E_c}{3V_c},
\end{equation}
\begin{equation}
\beta_c=\sqrt{ \frac{P_c}{\bar{\rho}_c c^2}},
\end{equation}
respectively, where $\bar{\rho}_c$ is the mean density of the full cocoon.

We consider a black hole with mass $M_{\rm BH}=10^6~M_\odot$ and $a=0.9$, and solve the above equations numerically for different density profiles and black hole parameters, {assuming that the jet is always conical.}
We stop the calculation when the ratio between the energy loss rate due to cocoon expansion and the jet power, $P_c\dot{V}_c/P_{\rm jet}$, reaches $\sim50\%$, since in principle our adopted formulae are only valid when this ratio is very small.
We assume an opening angle $\theta_0= 5^{\rm o}$, $\beta_j\approx1$.

\begin{figure}
\centering{
\subfigure{\includegraphics[width=0.45\textwidth]{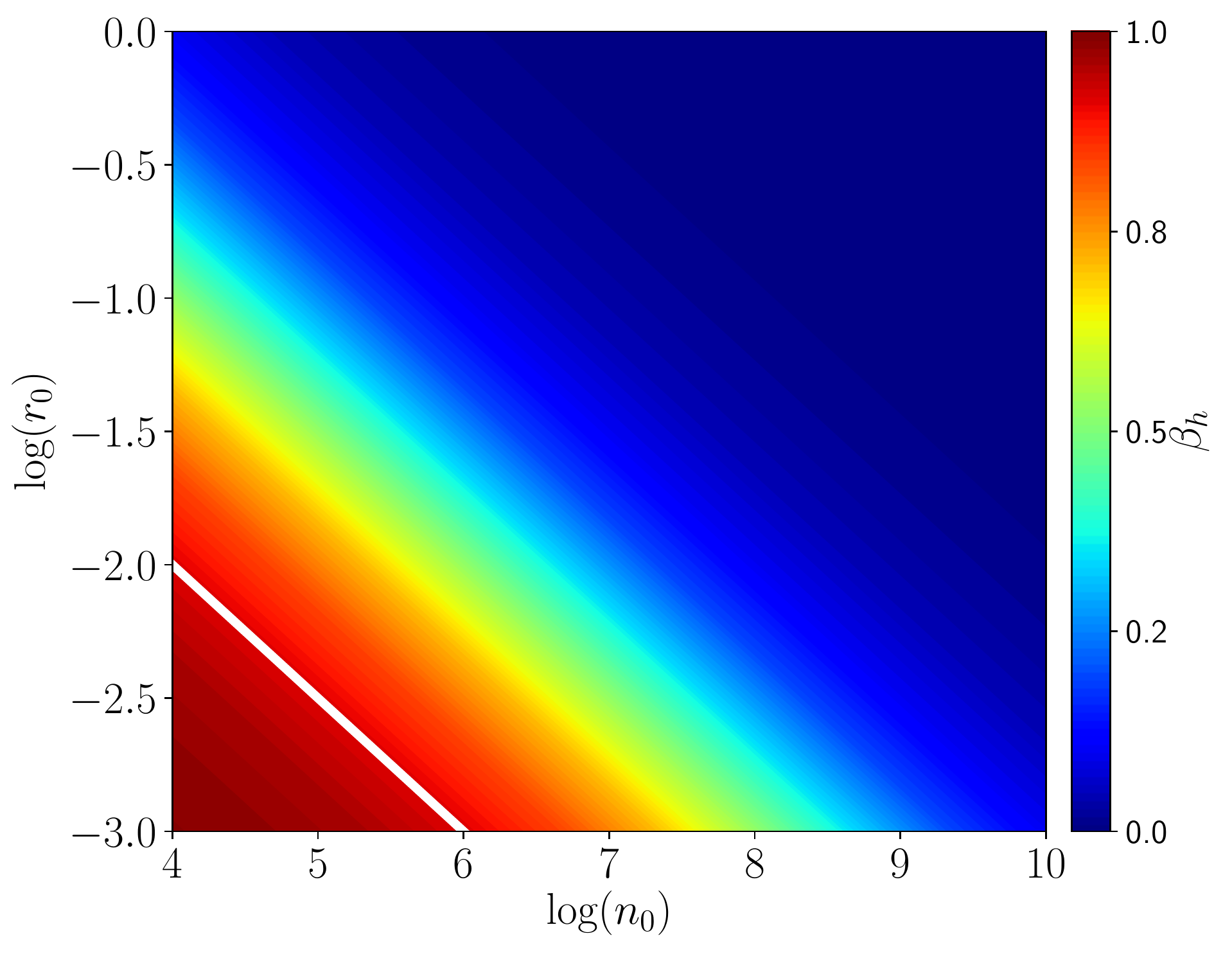}}
 \caption{
 {\it Top:}  The $\beta_h$ as a function of $n_0$ and $r_0$. Here $\beta_h$ is calculated either at the time when half of the jet power is deposited into the cocoon, or at the distance $>$ 10 pc from the central black hole. We take density profile slope $\alpha=2$. A white curve describes the relation $\log r_0=-0.5(\log n_0-4)-2$.
 \label{fig:beta_h}
}
}
\end{figure}

We show the  $\beta_h$ as a function of $n_0$ and $r_0$ in Fig. \ref{fig:beta_h}. Here $\beta_h$ is either at the point when $P_c\dot{V}_c/P_{\rm jet}=0.5$ (for jet that fails to break out the envelope), or at $r=10$ pc that is $\gg r_0$ (for jet that is considered to break out the envelope successfully). We find that, indeed, just in a small parameter space, say  roughly below the curve $\log r_0=-0.5(\log n_0-4)-2$ (the white curve), the jet is able to break out the envelope. However, we check that if the density profile is more steeper, for example the slope $\alpha=3$, then for more profiles (roughly below the curve $\log r_0=-1/3(\log n_0-4)-1$) the jet can break out the envelope.

Here we conclude that, if indeed the density profile has slope $\alpha\sim2$, as given by numerical simulations, lots of DCBHs may suffer from the free-free absorption by their envelops at SKA-low band unless the $N_{\rm H}\lesssim10^{22}$ cm$^{-2}$, and at SKA-mid band unless the $N_{\rm H}\lesssim10^{24}$ cm$^{-2}$.

\bsp	
\label{lastpage}
\end{document}